\documentclass[11pt,a4paper]{article}
\usepackage{jheppub}
\usepackage{slashed}
\usepackage{bbold}
\usepackage[czech,english]{babel}
\usepackage{amsmath,mathrsfs}
\usepackage{graphicx}
\usepackage{mathtools}
\DeclareMathOperator{\MyProd}{\scalebox{1.4}{$\mathrm{I\kern-0.2ex I}$}}

\preprint{YITP-SB-13-28}

\title{Localization of Supersymmetric Chern-Simons-Matter Theory on a Squashed $S^3$ with $SU(2)\times U(1)$ Isometry}

\author[a]{Jun Nian}

\affiliation[a]{C. N. Yang Institute for Theoretical Physics\\
	Stony Brook University\\
	Stony Brook, NY 11794-3840\\
	U.S.A.}

\emailAdd{jnian@insti.physics.sunysb.edu}

\abstract{Localization of supersymmetric $\mathcal{N}=2$ Chern-Simons-Matter theory on a squashed $S^3$ with $SU(2)\times U(1)$ isometry has been studied by different groups of authors. In this paper, we localize the theory on a squashed $S^3$ with $SU(2)\times U(1)$ isometry and a class of complex background. We see that certain kinds of shifts of the background gauge fields are crucial in obtaining nontrivial results, and the previously found results on this manifold can be incorporated in our results as special limits.}


\arxivnumber{}

\begin{document}
\maketitle
\flushbottom

\section{Introduction and Summary}

The idea of localization in supersymmetric theories orginated in Refs.~\cite{Witten-1, Witten-2}, and was further developed in Ref.~\cite{Nekrasov}. In recent years, it has revived due to the work of Pestun \cite{Pestun-1} and Kapustin, Willett, Yaakov \cite{Kapustin-1}. Inspired by these works, supersymmetric theories on curved spacetimes are being studied intensively \cite{Jaff, Jap-1, Jap-2, IY, Benini, Gomis-1, Gomis-2, 3Dblocks, Ellipsoid, S5, squashedS5}.

In the procedure of localizing a supersymmetric gauge theory on a curved manifold, a crucial step is to solve the Killing spinor equation, but usually this is a very formidable task. Recently, a new method \cite{Seiberg, Klare, 4D-I, 4D-II, 3D} has been proposed to deal with the supercharges on a 3D or 4D curved space. It has also been generalized to the 5D case \cite{Pan}. The idea of this new method can be applied to more general superspaces, as discussed in Ref.~\cite{Kuzenko}. Using this method, one can find the generalized Killing spinor equations on a curved manifold in a systematic way, and study how the partition functions of certain supersymmetric theories depend on the geometry of the manifold.

The basic idea is following. One starts with an off-shell minimal supergravity, which can be coupled to some other theories of interest, then sends the gravitino $\psi_\mu$ and its supersymmetry transformation $\delta \psi_\mu$ to zero. It will provide some Killing spinor equations with auxiliary fields. By studying the possible solutions to these Killing spinor equations for a specific curved space, one can construct a theory with rigid supersymmetry on that curved space \cite{Seiberg}. Given a curved manifold, one can also study the number of supercharges that can be defined on it \cite{4D-I, 4D-II, 3D}. To define supercharges, usually one has to turn on some background auxiliary fields, which will modify the theory and consequently appear in the final result of the partition function.

With this method in hand, one can study supersymmetric theories on a curved manifold more efficiently, for instance, 3D Chern-Simons-Matter theory. Superconformal Chern-Simons theories in flat spacetime were first studied by J.~Schwarz in Ref.~\cite{Schwarz-1}, where he discussed $\mathcal{N}=1$ and $\mathcal{N}=2$ superconformal Chern-Simons-Matter models in detail, while $\mathcal{N}=8$ models were discussed later in Ref.~\cite{Schwarz-2}. In Ref.~\cite{Kapustin-1} Kapustin, Willett and Yaakov generalized the $\mathcal{N}=2$ superconformal Chern-Simons-Matter theory from 3-dimensional flat Euclidean space to $S^3$, and calculated the partition function and the expectation value of the supersymmetric Wilson loop of the same theory on $S^3$. In Refs.~\cite{Jaff, Jap-1}, the idea of Ref.~\cite{Kapustin-1} was generalized and applied to the calculation of the partition function of the $\mathcal{N}=2$ supersymmetric Chern-Simons-Matter theory on $S^3$. In Refs.~\cite{Jap-2, IY} further generalizations were made, and the partition function of $\mathcal{N}=2$ supersymmetric Chern-Simons-Matter theory was calculated on different versions of squashed $S^3$.

Both Ref.~\cite{Jap-2} and Ref.~\cite{IY} have studied the localization of the theory on a squashed $S^3$ with $SU(2)\times U(1)$ isometry, but they chose different background gauge fields, hence their results are not the same. The gravity duals of these theories are studied in Ref.~\cite{two-param}. The different choices of the background have been discussed in Ref.~\cite{NUT} and more generally in Ref.~\cite{MS-new}. Using the method proposed in Ref.~\cite{3D}, Ref.~\cite{MS-new} has studied systematically the localization of $\mathcal{N}=2$ Chern-Simons-Matter theory on a squashed $S^3$ which is homotopic to $S^3$. As pointed out in Ref.~\cite{MS-new}, for a manifold with the same topology as $S^3$ the partition function is independent of the metric, instead it only depends on the choice of the background fields. Ref.~\cite{MS-new} gives the most general results for the real background fields. As a complement, we study the choice of complex background fields in this paper, which provides another possible interpolation between the results of Ref.~\cite{Jap-2} and Ref.~\cite{IY}.

We briefly summarize the main results of this paper. We studied the $\mathcal{N}=2$ Chern-Simons-Matter theory on a squashed $S^3$ with $SU(2)\times U(1)$ isometry \eqref{eq:squashedMetricInForms}:
    \begin{displaymath}
      ds^2 = \frac{\ell^2}{v^2} \mu^1 \mu^1 + \ell^2 \mu^2 \mu^2 + \ell^2 \mu^3 \mu^3\, ,
    \end{displaymath}
    where $v$ is a constant squashing parameter, $\ell$ is a constant length scale, and $\mu^m$ $(m = 1,\, 2,\, 3)$ denote the left-invariant 1-forms, which are discussed in Appendix B and C. To define supercharges on this manifold, some background auxiliary fields have to be turned on. The supercharges satisfy the generalized Killing spinor equation \eqref{eq:genKilling}:
  \begin{align}
    (\nabla_\mu - i A_\mu) \zeta & = -\frac{1}{2} H \gamma_\mu \zeta - i V_\mu \zeta - \frac{1}{2} \varepsilon_{\mu\nu\rho} V^\nu \gamma^\rho \zeta\, ,\nonumber\\
    (\nabla_\mu + i A_\mu) \widetilde{\zeta} & = -\frac{1}{2} H \gamma_\mu \widetilde{\zeta} + i V_\mu \widetilde{\zeta} + \frac{1}{2} \varepsilon_{\mu\nu\rho} V^\nu \gamma^\rho \widetilde{\zeta}\, .\nonumber
  \end{align}
Similar to Refs.~\cite{Jap-2, Ellipsoid}, we will first choose the Killing spinors $\zeta$ and $\widetilde{\zeta}$, and then solve for the auxiliary fields $A_\mu$, $V_\mu$ and $H$. In the left-invariant frame \eqref{eq:sqLeftinvFrame}
  \begin{displaymath}
    e^1 = \frac{\ell}{v} \mu^1\, ,\quad e^2 = \ell \mu^2\, ,\quad e^3 = \ell \mu^3\, ,
  \end{displaymath}
if we choose
  \begin{displaymath}
    \zeta_\alpha = \sqrt{s} \left(\begin{array}{c} 1 \\ 0 \end{array} \right)\, ,\quad \widetilde{\zeta}_\alpha = \frac{\Omega}{\sqrt{s}} \left(\begin{array}{c} 0 \\ 1 \end{array} \right)
  \end{displaymath}
  with
  \begin{displaymath}
    s = e^{i (\psi - \phi)}\, ,\quad \Omega = \frac{1}{2v}\, ,
  \end{displaymath}
  where $\psi$ and $\phi$ are coordinates in the Hopf fibration of $S^3$ \eqref{eq:metricHopf} discussed in Appendix C, then in general the auxiliary fields have the form \eqref{eq:sol} \eqref{eq:solH}:
  \begin{align}
    H & = \frac{i}{v\ell} + i\kappa\, ,\nonumber\\
    V_{1} & = \frac{2}{v\ell} + \kappa\, ,\quad V_{2} = V_{3} = 0\, ,\nonumber\\
    A_{1} & = \frac{v}{\ell} + \frac{2}{v\ell} + \frac{3\kappa}{2}\, ,\quad A_{2} = A_{3} = 0\, ,\nonumber
  \end{align}
where $\kappa$ can be some arbitrary constant. If we allow a further rotation of the Killing spinor \eqref{eq:rotateKilling}:
  \begin{displaymath}
    \zeta \rightarrow e^{i \gamma_1 \Theta} \zeta\, ,\quad \widetilde{\zeta} \rightarrow e^{-i \gamma_1 \Theta} \widetilde{\zeta}
  \end{displaymath}
with constant complex $\Theta$, then the most general background is given by \eqref{eq:genAux}:
    \begin{align}
      H & = \frac{i}{v\ell} + i \kappa\, \textrm{cos} (2\Theta) - (\frac{2}{v\ell} + \kappa) \textrm{sin} (2\Theta)\, , \nonumber\\
      V_1 & = (\frac{2}{v\ell} + \kappa) \textrm{cos} (2\Theta) + i \kappa\, \textrm{sin} (2\Theta)\, ,\quad V_2 = V_3 = 0\, , \nonumber\\
      A_1 & = \frac{v}{\ell} + (\frac{2}{v\ell} + \frac{3}{2} \kappa) \textrm{cos} (2\Theta) + (\frac{i}{v\ell} + \frac{3i}{2} \kappa) \textrm{sin} (2\Theta)\, ,\quad A_2 = A_3 = 0\, ,\nonumber
    \end{align}
    From the expressions above, it is clear that the effects of $\kappa$ and $\Theta$ are not the same, i.e., in general one cannot always make $\Theta = 0$ by choosing an appropriate $\kappa$. In other words, $\Theta$ describes some nontrivial shifts of the background auxiliary fields. Using the localization technique we calculated the partition function of the theory. The result is the following:
    \begin{equation}\label{eq:finalResult}
      Z = \frac{1}{|\mathcal{W}|} \int d^r \sigma Z_{\textrm{class}} \, Z_{\textrm{mat}}^{1-\textrm{loop}}\, Z_g^{1-\textrm{loop}}\, ,
    \end{equation}
    where $\sigma$ is the constant expectation value of the scalar in the vector multiplet which parametrizes the Coulomb branch of the theory, $|\mathcal{W}|$ denotes the order of the Weyl group, $r$ is the rank of the gauge group, and $Z_{\textrm{class}}$ is the classical value of the partition function, which is the product of the contributions \eqref{eq:Zclass-1} - \eqref{eq:Zclass-3} from $\mathscr{L}_{FI}$, $\mathscr{L}_{gg}$ and $\mathscr{L}_{gr}$ \eqref{eq:Lag-1} - \eqref{eq:Lag-3}, i.e.,
    \begin{align}
      Z_{\textrm{class}} = & \,\, \textrm{exp} \left( -\frac{4i \pi^2 \xi \ell^3}{v} H\, \textrm{Tr} (\sigma) \right) \cdot \textrm{exp} \left( \frac{i\pi k_{gg} \ell^3}{v} H\, \textrm{Tr} (\sigma^2) \right) \nonumber\\
      {} & \,\, \cdot \textrm{exp} \left( \frac{i \pi k_{gr} \ell^3}{2v} (H^2 + \frac{1}{2}R - V_\mu V^\mu)\, \textrm{Tr} (\sigma) \right)\, ,
    \end{align}
    where $\xi$, $k_{gg}$ and $k_{gr}$ denote the Fayet-Iliopoulos coupling, the Chern-Simons level and the gauge-$R$ Chern-Simons coupling respectively, and $R = \frac{8}{\ell^2} - \frac{2}{\ell^2 v^2}$ denotes the Ricci scalar of the squashed $S^3$. In Eq.~\eqref{eq:finalResult}, $Z_{\textrm{mat}}^{1-\textrm{loop}}$ and $Z_g^{1-\textrm{loop}}$ are the 1-loop determinants of the matter sector and the gauge sector respectively. They are given by Eq.~\eqref{eq:ZmatterSector} and Eq.~\eqref{eq:ZgaugeSector}:
    \begin{displaymath}
      Z_{\textrm{mat}}^{1-\textrm{loop}} = \prod_{\rho \in R} s_b \left(\frac{Q}{2} \left(\frac{z - q\rho(\sigma)}{\frac{v}{\ell_1}} - ir + i \right)\right)\, ,
    \end{displaymath}
    \begin{displaymath}
      Z_g^{1-\textrm{loop}} = \prod_{\alpha \in \Delta} s_b \left(\frac{Q}{2} \left(\frac{i\alpha(\sigma)}{\frac{v}{\ell_1}} - i \right) \right)\, ,
    \end{displaymath}
    where
    \begin{displaymath}
      Q \equiv b + b^{-1}\, ,\quad b \equiv \frac{1 - W}{\sqrt{1-W^2}} = \sqrt{\frac{1-W}{1+W}}\, ,\quad W \equiv \frac{\frac{v}{\ell} - \frac{v}{\ell_1}}{\frac{v}{\ell_1}}\, ,
    \end{displaymath}
    \begin{displaymath}
      \frac{v}{\ell_1} = A_1 - \frac{1}{2} V_1 + iH = \frac{v}{\ell} \left(1 - \frac{2i}{v^2}\, \textrm{sin}\Theta\, e^{-i\Theta} \right)\, ,
    \end{displaymath}
    while $\rho$ and $\alpha$ are the weights and the roots, $z$, $r$ and $q$ denote the central charge, the $R$-charge and the charge under the gauge group of the chiral multiplet repectively, and $s_b(x)$ is the double-sine function. We want to emphasize that $Z_{\textrm{class}}$ still has $\kappa$- and $\Theta$-dependence, while the 1-loop determinants $Z_{\textrm{mat}}^{1-\textrm{loop}}$ and $Z_g^{1-\textrm{loop}}$ contain only $\Theta$, i.e., the 1-loop determinants are independent of $\kappa$. Another important feature is that only $\Theta \neq 0$ can give the result with $b \neq 1$, which is like the result of the localization on a squashed $S^3$ with $U(1)\times U(1)$ isometry. This effect can be understood as follows: the background gauge fields twist the connection in the covariant derivative, and hence the naive $SU(2)$ degeneracy due to the original isometry $SU(2)\times U(1)$ is lifted, and the effective connection actually preserves only a $U(1)\times U(1)$ symmetry.

We want to emphasize that the background studied in this paper is not included in the analysis of Ref.~\cite{MS-new}, because the background fields in this paper are complex for non-zero $\Theta$, while Ref.~\cite{MS-new} limited their analysis for real background fields and conjectured that their analysis will go through for general complex backgrounds. This paper considers a special kind of complex backgrounds in detail, and it would be an interesting problem for future research to repeat the analysis for more general complex backgrounds.

This paper is organized as follows. In Section~2 we first discuss squashed $S^3$, then we recall some relevant formulae from Ref.~\cite{3D} and solve for Killing spinors and auxiliary fields on a squashed $S^3$ with $SU(2)\times U(1)$ isometry. In Section~3, $\mathcal{N}=2$ supersymmetric Chern-Simons-Matter theory is reviewed, and we make use of the Killing spinors and the auxiliary fields found in the previous section to localize the theory on the squashed $S^3$ with $SU(2)\times U(1)$ isometry. We end with a brief discussion in Section~4. Moreover, the convention of the paper is given in Appendix A, while Some metrics and frames on round $S^3$ as well as on squashed $S^3$ are summarized in Appendix B and C. The classical solutions to the BPS equations are given in Appendix D. In Appendix E, we discuss how to derive some important relations used in the paper.

\section{Killing Spinors and Auxiliary fields}

  \subsection{Review of Squashed $S^3$}
  In this section, we briefly discuss different squashed $S^3$'s and the corresponding Killing spinors that one can define on them. By squashed $S^3$ we mean the continuous deformation of the round $S^3$ metric by some parameters without changing the global topology. When these small parameters become zero, the metric of the squashed $S^3$ returns to the one of the round $S^3$.

  The metrics of squashed $S^3$ may have different isometry groups. As reviewed in Appendix B, the metric of round $S^3$ has $SU(2)_L \times SU(2)_R$ isometry. After squashing, the symmetry $SU(2)$ is reduced to some smaller group in the left-invariant frame or the right-invariant frame or both. Both Ref.~\cite{Jap-2} and Ref.~\cite{IY} have discussed squashed $S^3$ with isometry group smaller than $SU(2)_L \times SU(2)_R$. We adapt their expressions a little according to our convention.

  Ref.~\cite{Jap-2} introduced an example of squashed $S^3$ that preserves an $SU(2)_L \times U(1)_R$ isometry:
    \begin{equation}\label{eq:3DmetricSq-1}
      ds^2 = \tilde{\ell}\,^2\, \mu^1 \mu^1 + \ell^2 (\mu^2 \mu^2 + \mu^3 \mu^3)\, ,
    \end{equation}
    where in general the constant $\tilde{\ell}$ is different from the constant $\ell$, and $\mu^a\, (a = 1, 2, 3)$ are the left-invariant forms which are defined by
  \begin{equation}
    2 \mu^a T_a = g^{-1} dg\, ,\quad g \in SU(2)\, .
  \end{equation}
    In the frame
    \begin{equation}\label{eq:vielbeinHHL}
      (e^1, e^2, e^3) = (\tilde{\ell}\mu^1, \ell\mu^2, \ell\mu^3)
    \end{equation}
    the spin connections are
    \begin{equation}\label{eq:spinconnHHL}
      \omega^{23} = (2\tilde{\ell}\,^{-1} - f^{-1}) e^1\, ,\quad \omega^{31} = f^{-1} e^2\, ,\quad \omega^{12} = f^{-1} e^3\, ,
    \end{equation}
    where $f \equiv \ell^2 \tilde{\ell}\,^{-1}$. In this case, to define a Killing spinor, one has to turn on a background gauge field $V$. Then there can be two independent Killing spinors with opposite $R$-charges:
    \begin{equation}
      \nabla_m \epsilon = \frac{i}{2f} \gamma_m \epsilon + iV_m \epsilon\, ,\quad D_m \bar{\epsilon} = \frac{i}{2f} \gamma_m \bar{\epsilon} - iV_m \bar{\epsilon}\, ,
    \end{equation}
    where
    \begin{equation}
      \nabla_m \epsilon \equiv \partial_m \epsilon + \frac{1}{4} \gamma_{ab} \omega_m^{ab} \epsilon \, ,\quad \nabla_m \bar{\epsilon} \equiv \partial_m \bar{\epsilon} + \frac{1}{4} \gamma_{mn} \omega_m^{mn} \bar{\epsilon}\, ,
    \end{equation}
    and
    \begin{equation}\label{eq:bgd-HHL}
      V_m = e_m^1 \left(\frac{1}{\tilde{\ell}} - \frac{1}{f} \right)\, ,\qquad \epsilon = \left(\begin{array}{cc}1\\0 \end{array}\right)\, ,\qquad \bar{\epsilon} = \left(\begin{array}{cc}0\\1 \end{array}\right) .
    \end{equation}

  The same metric with $SU(2)_L \times U(1)_R$ isometry was also considered in Ref.~\cite{IY}:
  \begin{equation}\label{eq:secondmetric}
    ds^2 = \ell^2 \left(\frac{1}{v^2} \mu^1 \mu^1 + \mu^2 \mu^2 + \mu^3 \mu^3 \right)\, ,
  \end{equation}
  where $\ell$ is a constant with dimension of length, and $v$ is the constant squashing parameter. This metric is related to the previous case of squashed $S^3$ in the following way:
  \begin{equation}
    \frac{\ell}{v} = \tilde{\ell}\, .
  \end{equation}
  The vielbeins and the spin connections are the same as in the previous case, i.e., they are still given by Eq.~\eqref{eq:vielbeinHHL} and Eq.~\eqref{eq:spinconnHHL} respectively. However, Ref.~\cite{IY} chose a different background gauge field $V_m$, and the Killing spinor equations are
  \begin{equation}
    \nabla_m \epsilon = -\frac{i}{2v\ell} \gamma_m \epsilon + \frac{u}{v\ell} V^n \gamma_{mn} \epsilon\, ,\quad \nabla_m \bar{\epsilon} = -\frac{i}{2v\ell} \gamma_m \bar{\epsilon} - \frac{u}{v\ell} V^n \gamma_{mn} \bar{\epsilon}\, ,
  \end{equation}
  where again
    \begin{equation}
      \nabla_m \epsilon \equiv \partial_m \epsilon + \frac{1}{4} \gamma_{ab} \omega_m^{ab} \epsilon\, ,\quad \nabla_m \bar{\epsilon} \equiv \partial_m \bar{\epsilon} + \frac{1}{4} \gamma_{ab} \omega_m^{ab} \bar{\epsilon}\, ,
    \end{equation}
  and $u$ is defined by
  \begin{equation}\label{eq:relvu}
    v^2 = 1 + u^2\, ,
  \end{equation}
  while the background gauge field is given by
  \begin{equation}\label{eq:bgd-IY}
    V^m = e_1^m\, .
  \end{equation}
  The Killing spinors in this case have the solution:
  \begin{equation}
    \epsilon = e^{\theta \frac{\sigma_3}{2i}} g^{-1} \epsilon_0\, ,\quad \bar{\epsilon} = e^{-\theta \frac{\sigma_3}{2i}} g^{-1} \bar{\epsilon}_0\, ,
  \end{equation}
  where $\epsilon_0$ and $\bar{\epsilon}_0$ are arbitrary constant spinors, and the angle $\theta$ is given by
  \begin{equation}
    e^{i\theta} = \frac{1 + i u}{v}\, .
  \end{equation}

    Actually there is another example of squashed $S^3$ discussed in Ref.~\cite{Jap-2}:
    \begin{equation}\label{eq:3DmetricSq-2-1}
      ds^2 = \ell^2 (dx_0\,^2 + dx_1\,^2) + \tilde{\ell}\,^2 (dx_2\,^2 + dx_3\,^2)\, .
    \end{equation}
    This metric preserves an $U(1)\times U(1)$ isometry. Transforming the coordinates $(x_0, x_1, x_2, x_3)$ to $(\textrm{cos}\theta\, \textrm{cos}\varphi, \textrm{cos}\theta\, \textrm{sin}\varphi, \textrm{sin}\theta\, \textrm{cos}\chi, \textrm{sin}\theta\, \textrm{sin}\chi)$, we can rewrite the metric as:
    \begin{equation}\label{eq:3DmetricSq-2-2}
      ds^2 = f(\theta)^2 d\theta^2 + \ell^2 \textrm{cos}^2 \theta\, d\varphi^2 + \tilde{\ell}\,^2 \textrm{sin}^2 \theta\, d\chi^2\, ,
    \end{equation}
    where
    \begin{equation}
      f(\theta) \equiv \sqrt{\ell^2 \textrm{sin}^2 \theta + \tilde{\ell}\,^2 \textrm{cos}^2 \theta}\, .
    \end{equation}
    It is discussed in Ref.~\cite{Jap-2} in great detail. Since we focus on the one with $SU(2)\times U(1)$ isometry, we will not consider this case in the following.

  \subsection{Some Relevant Formulae}\label{sec:method}

  In this section we recall some relevant formulae from Ref.~\cite{3D}, which describes a set of generalized 3D Killing spinor equations. By adding a flat direction to a 3D manifold one obtains a 4D manifold, so in principle the 4D formalisms introduced in Refs.~\cite{4D-I, 4D-II} can also be applied to a squashed $S^3$. We will focus on the 3D formalism \cite{3D} in the following. The generalized Killing spinor equations discussed in Ref.~\cite{3D} are:
  \begin{align}\label{eq:genKilling}
    (\nabla_\mu - i A_\mu) \zeta & = -\frac{1}{2} H \gamma_\mu \zeta - i V_\mu \zeta - \frac{1}{2} \varepsilon_{\mu\nu\rho} V^\nu \gamma^\rho \zeta\, ,\\
    (\nabla_\mu + i A_\mu) \widetilde{\zeta} & = -\frac{1}{2} H \gamma_\mu \widetilde{\zeta} + i V_\mu \widetilde{\zeta} + \frac{1}{2} \varepsilon_{\mu\nu\rho} V^\nu \gamma^\rho \widetilde{\zeta}\, .
  \end{align}
  The Killing spinor equations in Refs.~\cite{Jap-2, IY} can be viewed as these generalized Killing spinor equations with special choices of the auxiliary fields $A_m$, $V_m$ and $H$. If we choose the metric \eqref{eq:squashedMetricInForms} in the left-invariant frame \eqref{eq:sqLeftinvFrame}:
  \begin{equation}
    ds^2 = \frac{\ell^2}{v^2} \mu^1 \mu^1 + \ell^2 \mu^2 \mu^2 + \ell^2 \mu^3 \mu^3\, ,
  \end{equation}
  \begin{displaymath}
    e^1 = \frac{\ell}{v} \mu^1\, ,\quad e^2 = \ell \mu^2\, ,\quad e^3 = \ell \mu^3\, ,
  \end{displaymath}
  then the choice in Ref.~\cite{Jap-2} is
  \begin{align}\label{eq:auxChoice-1}
    A_1 & = \frac{v}{\ell} - \frac{1}{v\ell}\, ,\quad A_2 = A_3 = 0\, ,\nonumber\\
    V_m & = 0\, ,\quad (m = 1,\, 2,\, 3)\nonumber\\
    H & = -\frac{i}{v\ell}\, ,
  \end{align}
  while Ref.~\cite{IY} chose
  \begin{align}\label{eq:auxChoice-2}
    A_1 & = V_1 = -\frac{2 i u}{v\ell}\, , \nonumber\\
    A_2 & = A_3 = V_2 = V_3 = 0\, ,\nonumber\\
    H & = \frac{i}{v\ell}\, ,
  \end{align}
  where $v$ is the constant squashing parameter, $u \equiv \sqrt{v^2 - 1}$, and $\ell$ denotes the length scale.

  We make use of the formalism described in Ref.~\cite{3D} to solve for the Killing spinors and the background auxiliary fields on the squashed $S^3$ discussed above. We expect that in some limits the results of Refs.~\cite{Jap-2, IY} can be reproduced within the framework of Ref.~\cite{3D}. The following sketch illustrates the path of calculations:
  \begin{equation}
    \textrm{Define}\quad K_m \equiv \zeta \gamma_m \widetilde{\zeta}\, ,\quad \eta_m \equiv \Omega^{-1} K_m\, ,\quad \Phi^m\,_n \equiv \varepsilon^m\,_{np} \eta^p\, ,\quad P_m \equiv \zeta \gamma_m \zeta
  \end{equation}
  \begin{displaymath}
    \Downarrow
  \end{displaymath}
  \begin{equation}\label{eq:Def-s}
    \textrm{Define}\quad p \equiv P_{\bar{z}}\, ,\quad s \equiv \frac{1}{\sqrt{2}} p g^{-\frac{1}{4}} \sqrt{\Omega}\, ,\quad W_m \equiv -\frac{1}{4} \eta_m \varepsilon^{npq} \eta_n \partial_p \eta_q
  \end{equation}
  \begin{equation}
    V^m = \varepsilon^{mnp} \partial_n \eta_p + \kappa \eta^m
  \end{equation}
  \begin{equation}
    H = -\frac{1}{2} \nabla_m \eta^m + \frac{i}{2} \varepsilon^{mnp} \eta_m \partial_n \eta_p + i \kappa
  \end{equation}
  \begin{displaymath}
    \Downarrow
  \end{displaymath}
  \begin{equation}
    A_m = \frac{1}{8} \Phi_m\,^n \partial_n \textrm{log} g - \frac{i}{2} \partial_m \textrm{log} s + \frac{1}{2} (2 \delta_m\,^n - i \Phi_m\,^n) V_n - \frac{i}{2} \eta_m H + W_m + \frac{3}{2} \kappa \eta_m
  \end{equation}
  Finally, we obtain the auxiliary fields $V_m$, $H$ and $A_m$. We should emphasize that the factor $g$ appearing in the definition of $s$ is the absolute value of the determinant of the metric with the form
  \begin{equation}\label{eq:metricForm}
    ds^2 = \Omega^2 (d\psi + a dz + \bar{a} d\bar{z})^2 + c^2 dz d\bar{z}\, ,
  \end{equation}
  and $p$ is defined as the $\bar{z}$-component of $P_m$ in this coordinate system.

  As pointed out in Ref.~\cite{3D}, for the Killing spinors and the auxiliary fields satisfying the Killing spinor equations \eqref{eq:genKilling}, one can shift the auxiliary fields while preserving the same Killing spinors:
  \begin{align}
    V^\mu & \rightarrow V^\mu + \kappa \eta^\mu\, , \nonumber\\
    H & \rightarrow H + i\kappa\, ,\nonumber\\
    A^\mu & \rightarrow A^\mu + \frac{3}{2} \kappa \eta^\mu\, ,
  \end{align}
  where $\kappa$ satisfies
  \begin{equation}
    K^\mu \partial_\mu \kappa = 0\, .
  \end{equation}
  It means that after obtaining a set of solutions of the auxiliary fields, one can always shift them to obtain new solutions without changing the Killing spinors, and the new auxiliary fields and the Killing spinors formally satisfy the same Killing spinor equations as before.

  \subsection{Solving for Killing Spinors and Auxiliary Fields}

  Following the path which is summarized in the previous subsection, we solve for the Killing spinors and the auxiliary fields for the squashed $S^3$ with $SU(2)\times U(1)$ isometry. Starting from the metric in the left-invariant frame \eqref{eq:squashedMetricInForms}
  \begin{displaymath}
    ds^2 = \frac{\ell^2}{v^2} \mu^1 \mu^1 + \ell^2 \mu^2 \mu^2 + \ell^2 \mu^3 \mu^3\, ,
  \end{displaymath}
  as discussed in Appendix C, we can first rewrite it into the form of Eq.~\eqref{eq:metric-complex}:
  \begin{displaymath}
    ds^2 = \frac{1}{4v^2} (d\psi + a dz + \bar{a} d \bar{z})^2 + c^2 dz\, d\bar{z}\, ,
  \end{displaymath}
  where we omit the length scale $\ell$ for simplicity, and consequently it will be omitted in the auxiliary fields, but we will bring it back in the end. Comparing this expression with Eq.~\eqref{eq:metricForm}, we can read off
  \begin{equation}
    \Omega = \frac{1}{2v}\, .
  \end{equation}
  We choose the Killing spinors to be
  \begin{equation}\label{eq:KillingSp}
    \zeta_\alpha = \sqrt{s} \left(\begin{array}{c} 1 \\ 0 \end{array} \right)\, ,\quad \widetilde{\zeta}_\alpha = \frac{\Omega}{\sqrt{s}} \left(\begin{array}{c} 0 \\ 1 \end{array} \right) = \frac{1}{2 v \sqrt{s}} \left(\begin{array}{c} 0 \\ 1 \end{array} \right)\, ,
  \end{equation}
  and use the matrix
  \begin{equation}
    \varepsilon^{\alpha\beta} = \left(\begin{array}{cc}
                                        0 & 1\\
                                        -1 & 0
                                      \end{array} \right)\, .
  \end{equation}
  to raise the indices of $\zeta_\alpha$ and $\widetilde{\zeta}_\alpha$:
  \begin{equation}\label{eq:KillingSpUp}
    \zeta^\alpha = \sqrt{s} \left(\begin{array}{c} 0 \\ -1 \end{array} \right) \, ,\quad \widetilde{\zeta}^\alpha = \frac{1}{2v\sqrt{s}} \left(\begin{array}{c} 1 \\ 0 \end{array} \right)\, .
  \end{equation}
  Next, we calculate $K_m$ in the working frame $(\hat{e}_1,\, \hat{e}_2,\, \hat{e}_3)$ \eqref{eq:workingFrame} \eqref{eq:workingFrame-2}. For practical reason, we will mainly work in this frame. Only in the end, we will bring the final results into the left-invariant frame \eqref{eq:sqLeftinvFrame}. In the following, without special mentioning the index $m = 1,\, 2,\, 3$ denotes the frame $(\hat{e}_1,\, \hat{e}_2,\, \hat{e}_3)$ \eqref{eq:workingFrame} \eqref{eq:workingFrame-2}.
  \begin{align}
    K_1 & = \zeta \gamma_1 \widetilde{\zeta} = \frac{1}{2v}\, ,\nonumber\\
    K_2 & = \zeta \gamma_2 \widetilde{\zeta} = 0\, ,\nonumber\\
    K_3 & = \zeta \gamma_3 \widetilde{\zeta} = 0\, .
  \end{align}
  In the coordinates $(X,\, Y,\, \psi)$ \eqref{eq:stereoCoord}, $K_m$ are given by
  \begin{align}
    K_X & = -\frac{1}{4 v^2} \cdot \frac{X^2+Y^2-1}{X^2+Y^2+1} \cdot \frac{Y}{X^2+Y^2}\, ,\nonumber\\
    K_Y & = \frac{1}{4 v^2} \cdot \frac{X^2+Y^2-1}{X^2+Y^2+1} \cdot \frac{X}{X^2+Y^2}\, ,\nonumber\\
    K_\psi & = \frac{1}{4 v^2}\, ,
  \end{align}
  while $K^m$ have a relatively simple form:
  \begin{equation}
    K^X = 0\, ,\quad K^Y = 0\, ,\quad K^\psi = 1\, .
  \end{equation}
  They satisfy
  \begin{equation}
    K^m K_m = \frac{1}{4v^2} = \Omega^2\, .
  \end{equation}
  $\eta_m$ can be obtained immediately
  \begin{equation}
    \eta_m = \frac{1}{\Omega} K_m = 2 v K_m\, ,
  \end{equation}
  i.e.,
  \begin{equation}
    \eta_1 = 1\, ,\quad \eta_2 = \eta_3 = 0\, .
  \end{equation}
  Then
  \begin{equation}
    \Phi^m\,_n \equiv \varepsilon^m\,_{np} \eta^p = \varepsilon^m\,_{n1} \eta^1 = \varepsilon^m\,_{n1}\, .
  \end{equation}
  Similarly,
  \begin{align}
    P_1 & =  \zeta \gamma_1 \zeta = 0\, ,\nonumber\\
    P_2 & =  \zeta \gamma_2 \zeta = s\, ,\nonumber\\
    P_3 & =  \zeta \gamma_3 \zeta = -i s\, ,\nonumber\\
  \end{align}
  and
  \begin{align}
    P_X & = \frac{s}{1+X^2+Y^2}\, ,\nonumber\\
    P_Y & = \frac{-i s}{1+X^2+Y^2}\, ,\nonumber\\
    P_\psi & = 0\, .
  \end{align}
  Since
  \begin{equation}
    P_z dz + P_{\bar{z}} d\bar{z} = P_z (dX + i dY) + P_{\bar{z}} (dX - i dY) = (P_z + P_{\bar{z}}) dX + i (P_z - P_{\bar{z}}) dY\, ,
  \end{equation}
  there is
  \begin{equation}
    P_X = P_z + P_{\bar{z}}\, ,\quad P_Y = i (P_z - P_{\bar{z}})\, ,
  \end{equation}
  or equivalently
  \begin{equation}
    P_z = \frac{1}{2} (P_X - i P_Y)\, ,\quad P_{\bar{z}} = \frac{1}{2} (P_X + i P_Y)\, .
  \end{equation}
  Then in this case
  \begin{equation}
    p \equiv P_{\bar{z}} = \frac{1}{2} (P_X + i P_Y) = \frac{s}{1+X^2+Y^2}\, .
  \end{equation}
  Plugging it into the definition of $s$ given by Eq.~\eqref{eq:Def-s}, we obtain
  \begin{equation}
    \frac{1}{\sqrt{2}} p g^{-\frac{1}{4}} \sqrt{\Omega} = s\, .
  \end{equation}
  Hence, the results are consistent, and we still have the freedom to choose the function $s$.

  It is straightforward to calculate $V_m$, $H$ and $W_m$:
  \begin{equation}
    V^1 = \frac{2}{v} + \kappa\, ,\quad V^2 = 0\, ,\quad V^3 = 0\, ,
  \end{equation}
  \begin{equation}
    H = \frac{i}{v} + i \kappa\, ,
  \end{equation}
  \begin{equation}
    W_1 = -\frac{1}{2v}\, ,\quad W_2 = 0\, ,\quad W_3 = 0\, .
  \end{equation}
  To calculate $A_m$, we first calculate $\hat{A}_m$:
  \begin{equation}
    \hat{A}_m \equiv \frac{1}{8} \Phi_m\,^n \partial_n \textrm{log} g - \frac{i}{2} \partial_m \textrm{log} s\, ,
  \end{equation}
  which is valid only in the coordinates $(z,\, \bar{z},\, \psi)$ \eqref{eq:complexCoord}. Using the definition of $\Phi^m\,_n$ we obtain
  \begin{equation}
    \Phi^z\,_z = -i\, ,\quad \Phi^{\bar{z}}\,_{\bar{z}} = i\, ,\quad \Phi^z\,_{\bar{z}} = \Phi^{\bar{z}}\,_z = 0\, ,
  \end{equation}
  and
  \begin{equation}
    \Phi_n\,^m = - \Phi^m\,_n\, .
  \end{equation}
  Therefore,
  \begin{align}
    \hat{A}_z & = -\frac{i}{2} \cdot \frac{\bar{z}}{1 + z \bar{z}} - \frac{i}{2} \partial_z \textrm{log} s\, ,\nonumber\\
    \hat{A}_{\bar{z}} & = \frac{i}{2} \cdot \frac{z}{1 + z \bar{z}} - \frac{i}{2} \partial_{\bar{z}} \textrm{log} s\, ,\nonumber\\
    \hat{A}_\psi & = -\frac{i}{2} \partial_\psi \textrm{log} s\, .
  \end{align}
  Then we obtain
  \begin{align}
    \hat{A}_1 & = \hat{A}_\psi e_1^\psi = -i v \partial_\psi \textrm{log} s\, ,\nonumber\\
    \hat{A}_2 & = \hat{A}_z e_2^z + \hat{A}_{\bar{z}} e_2^{\bar{z}} + \hat{A}_\psi e_2^\psi\nonumber\\
    {} & = \frac{i}{2} (z - \bar{z}) \left[1 + \frac{i}{2} (\partial_\psi \textrm{log} s) \frac{z\bar{z} - 1}{z \bar{z}}  \right] - \frac{i}{2} (1+z\bar{z}) (\partial_z \textrm{log} s + \partial_{\bar{z}} \textrm{log} s)\nonumber\\
    {} & = \frac{i}{2} (z - \bar{z}) \left[1 + \frac{i}{2} (\partial_\psi \textrm{log} s) \frac{z\bar{z} - 1}{z \bar{z}}  \right] - \frac{i}{2} (1+z\bar{z}) \partial_X \textrm{log} s \, ,\nonumber\\
    \hat{A}_3 & = \frac{1}{2} (z + \bar{z}) \left[1 + \frac{i}{2} (\partial_\psi \textrm{log} s) \frac{z\bar{z} - 1}{z \bar{z}}  \right] + \frac{1}{2} (1+z\bar{z}) (\partial_z \textrm{log} s - \partial_{\bar{z}} \textrm{log} s)\nonumber\\
    {} & = \frac{1}{2} (z + \bar{z}) \left[1 + \frac{i}{2} (\partial_\psi \textrm{log} s) \frac{z\bar{z} - 1}{z \bar{z}}  \right] - \frac{i}{2} (1+z\bar{z}) \partial_Y \textrm{log} s \, ,
  \end{align}
  and
  \begin{align}
    A_1 & = \hat{A}_1 + V_1 - \frac{i}{2} \Phi_1\,^n V_n - \frac{i}{2} \eta_1 H + W_1 + \frac{3}{2} \kappa \eta_1 \nonumber\\
    {} & = -i v \partial_\psi \textrm{log} s + \frac{2}{v} - \frac{i}{2} \frac{i}{v} - \frac{1}{2v} + \frac{3}{2} \kappa \nonumber\\
    {} & = -i v \partial_\psi \textrm{log} s + \frac{2}{v} + \frac{3}{2} \kappa \nonumber\\
    A_2 & = \hat{A}_2 + V_2 - \frac{i}{2} \Phi_2\,^n V_n - \frac{i}{2} \eta_2 H + W_2 + \frac{3}{2} \kappa \eta_2 \nonumber\\
    {} & = \frac{i}{2} (z - \bar{z}) \left[1 + \frac{i}{2} (\partial_\psi \textrm{log} s) \frac{z\bar{z} - 1}{z \bar{z}}  \right] - \frac{i}{2} (1+z\bar{z}) \partial_X \textrm{log} s \, ,\nonumber\\
    A_3 & = \hat{A}_3 + V_3 - \frac{i}{2} \Phi_3\,^n V_n - \frac{i}{2} \eta_3 H + W_3 + \frac{3}{2} \kappa \eta_3 \nonumber\\
    {} & = \frac{1}{2} (z + \bar{z}) \left[1 + \frac{i}{2} (\partial_\psi \textrm{log} s) \frac{z\bar{z} - 1}{z \bar{z}}  \right] - \frac{i}{2} (1+z\bar{z}) \partial_Y \textrm{log} s \, ,
  \end{align}
  where
  \begin{displaymath}
    V_1 = \frac{2}{v}\, ,\quad H = \frac{i}{v}\, .
  \end{displaymath}

  Our working frame $(\hat{e}_1,\, \hat{e}_2,\, \hat{e}_3)$ \eqref{eq:workingFrame} is not the left-invariant frame $(e_1,\, e_2,\, e_3)$ \eqref{eq:sqLeftinvFrame}. To transform between different frames, it is convenient to first consider the $\theta$-, $\phi$- and $\psi$-component of the fields, because different frames all have the same form of the metric \eqref{eq:metric-thetaphipsi}. Let us first calculate $V_\mu$ and $A_\mu$ $(\mu = \theta,\, \phi,\, \psi)$, then transform them into other frames. $V_\mu$ can be obtained very easily:
  \begin{align}
    V_\theta & = V_1 \hat{e}^1_\theta = 0\, ,\nonumber\\
    V_\phi & = V_1 \hat{e}^1_\phi = \left(\frac{2}{v} + \kappa\right) \frac{1}{2v} \textrm{cos} \theta\, ,\nonumber\\
    V_\psi & = V_1 \hat{e}^1_\psi = \left(\frac{2}{v} + \kappa\right) \frac{1}{2v}\, .
  \end{align}
  $A_\mu$ can also be calculated:
  \begin{align}
    A_\theta & = \frac{i}{2 \textrm{sin}\theta} (X \partial_X \textrm{log} s + Y \partial_Y \textrm{log} s)\, ,\nonumber\\
    A_\phi & = (\frac{1}{2} + \frac{1}{v^2} + \frac{3\kappa}{4v}) \textrm{cos}\theta + \frac{1}{2} + \frac{i}{2} (Y \partial_X \textrm{log} s - X \partial_Y \textrm{log} s)\, ,\nonumber\\
    A_\psi & = -\frac{i}{2} \partial_\psi \textrm{log} s + \frac{1}{v^2} + \frac{3\kappa}{4v}\, ,
  \end{align}
  where $\kappa$ should satisfy
  \begin{equation}
    K^m \partial_m \kappa = 0\, .
  \end{equation}
  Obeying this constraint it seems that we can choose any $\kappa$ and $s$, but as in Refs.~\cite{Jap-2, IY} we do not want to turn on the $2$- and $3$-component of $V_m$ and $A_m$ in the left-invariant frame \eqref{eq:sqLeftinvFrame}, because the deformation of the metric happens only in the $1$-direction. For this reason we always set $A_\theta = 0$, because it is contributed only from $A_2$ and $A_3$ in the left-invariant frame \eqref{eq:sqLeftinvFrame}. Hence, if $A_2 = A_3 = 0$, $A_\theta$ should also vanish.
  \begin{equation}
    A_\theta = 0\quad \Rightarrow \quad X \partial_X \textrm{log} s + Y \partial_Y \textrm{log} s = 0\, .
  \end{equation}
  The solution to this equation is still quite general, which is
  \begin{equation}
    \textrm{log} s = f(\psi) \cdot g\left(\frac{X}{Y}\right)\, ,
  \end{equation}
  where $f(x)$ and $g(x)$ can be any regular functions. A possible solution to $A_2=0$ and $A_3=0$ is
  \begin{equation}
    \textrm{log} s = -i\, \textrm{arctan}\left(\frac{Y}{X}\right) + i\psi \quad \Rightarrow \quad s = e^{i (\psi - \phi)}\, .
  \end{equation}
  With this choice there are
  \begin{align}
    A_\theta & = 0\, ,\nonumber\\
    A_\phi & = (\frac{1}{2} + \frac{1}{v^2} + \frac{3\kappa}{4v}) \textrm{cos}\theta\, ,\nonumber\\
    A_\psi & = \frac{1}{2} + \frac{1}{v^2} + \frac{3\kappa}{4v}\, .
  \end{align}
  Transforming $V_m$ and $A_m$ given above into the left-invariant frame \eqref{eq:sqLeftinvFrame}, we obtain:
  \begin{align}\label{eq:sol}
    V_1 & = \frac{2}{v} + \kappa\, ,\nonumber\\
    V_2 & = V_3 = 0\, ,\nonumber\\
    A_1 & = v + \frac{2}{v} + \frac{3\kappa}{2}\, ,\nonumber\\
    A_2 & = A_3 = 0\, ,
  \end{align}
  while $H$ has the form:
  \begin{equation}\label{eq:solH}
    H = \frac{i}{v} + i\kappa\, .
  \end{equation}

  Now we can try to reproduce the choices of the auxiliary fields in Refs.~\cite{Jap-2, IY} using our results \eqref{eq:sol} \eqref{eq:solH} obtained above. Ref.~\cite{Jap-2} made a special choice
  \begin{equation}
    \kappa = -\frac{2}{v}\, ,
  \end{equation}
  hence setting $\ell = 1$ they had
  \begin{equation}
    A_{1} = v - \frac{1}{v}\, ,
  \end{equation}
  or equivalently
  \begin{equation}
    A_{\phi} = (\frac{1}{2} - \frac{1}{2v^2}) \textrm{cos} \theta\, ,\quad A_{\psi} = \frac{1}{2} - \frac{1}{2v^2}\, ,
  \end{equation}
  and all the other components of $V_m$ and $A_m$ vanish.

  To reproduce the results in Ref.~\cite{IY}, things are a little involved, because there are no obvious solutions which can satisfy the conditions
  \begin{displaymath}
    H = \frac{i}{v}\, ,\quad V_{1} = A_{1} =  -\frac{2 i u}{v}\, ,
  \end{displaymath}
  where $u \equiv \sqrt{v^2 - 1}$. We have to consider other freedom in the solution. The auxiliary fields are given by Eqs.~\eqref{eq:sol}-\eqref{eq:solH} and the Killing spinor is given by Eq.~\eqref{eq:KillingSp}:
  \begin{displaymath}
    \zeta_\alpha = \sqrt{s} \left(\begin{array}{c} 1 \\ 0 \end{array} \right)\, ,\quad \widetilde{\zeta}_\alpha = \frac{\Omega}{\sqrt{s}} \left(\begin{array}{c} 0 \\ 1 \end{array} \right)\, ,
  \end{displaymath}
  where
  \begin{displaymath}
    s = e^{i (\psi - \phi)}\, ,\quad \Omega = \frac{1}{2v}\, .
  \end{displaymath}
  Suppose that we have obtained a set of solutions to the Killing spinor equations \eqref{eq:genKilling}, i.e. Killing spinors and corresponding auxiliary fields. Then we can rotate the Killing spinors by a constant angle $\Theta$ in the following way:
  \begin{equation}\label{eq:rotateKilling}
    \zeta \rightarrow e^{i \gamma_1 \Theta} \zeta\, ,\quad \widetilde{\zeta} \rightarrow e^{-i \gamma_1 \Theta} \widetilde{\zeta}\, .
  \end{equation}
  In order that the Killing spinor equations \eqref{eq:genKilling} still hold, the auxiliary fields have to be shifted correspondingly:
  \begin{align}\label{eq:genAux}
    H \rightarrow H' & = H\, \textrm{cos} (2\Theta) - V_1\, \textrm{sin} (2\Theta) - i\, \omega_2\,^{31} (1 - \textrm{cos} (2\Theta)) \nonumber\\
    {} & = \frac{i}{v} + i \kappa\, \textrm{cos} (2\Theta) - (\frac{2}{v} + \kappa) \textrm{sin} (2\Theta)\, , \nonumber\\
    V_1 \rightarrow V'_1 & = V_1 \textrm{cos}\, (2\Theta) + H\, \textrm{sin} (2\Theta) + i\, \omega_2\,^{31} \textrm{sin} (2\Theta)\nonumber\\
    {} & = (\frac{2}{v} + \kappa) \textrm{cos} (2\Theta) + i \kappa\, \textrm{sin} (2\Theta)\, , \nonumber\\
    A_1 \rightarrow A'_1 & = A_1 - \frac{i}{2} (H' - H) + (V'_1 - V_1) \nonumber\\
    {} & = v + (\frac{2}{v} + \frac{3}{2} \kappa) \textrm{cos} (2\Theta) + (\frac{i}{v} + \frac{3i}{2} \kappa) \textrm{sin} (2\Theta)\, .
  \end{align}
  where $\omega_2\,^{31}$ is one of the spin connections \eqref{eq:leftSpConn} in the left-invariant frame \eqref{eq:sqLeftinvFrame}. From the expressions above, we see that the effects of $\kappa$ and $\Theta$ are not the same, i.e., in general one cannot always make $\Theta = 0$ by choosing an appropriate $\kappa$. So we still have the freedom to choose $\Theta$ and $\kappa$, where $\Theta$ is in general complex. Moreover, until now we have omitted the length scale $\ell$, and actually rescaling $\ell$ also leaves the Killing spinor equations \eqref{eq:genKilling} invariant, hence it is a symmetry. Therefore, by choosing $\Theta$, $\kappa$ and $\ell$ we can make the conditions required by Ref.~\cite{IY} valid simultaneously:
  \begin{align}
    V'_1 & = A'_1\, , \\
    \frac{V'_1}{H'} & = - 2 u\, , \\
    H' & = \frac{i}{v \ell_0}\, ,
  \end{align}
  where $u \equiv \sqrt{v^2 - 1}$. The constraints above have a solution:
  \begin{align}
    \kappa & = -5 v + 7 v^3 - 4 v^5\, , \label{eq:IYsol-1}\\
    \ell & = \ell_0 (1 - 2 v^2)\, , \label{eq:IYsol-2}\\
    \Theta & = \textrm{arctan} \left(\frac{2i + i \kappa v - 2 \sqrt{-1 - \kappa v - v^4}}{\kappa v - 2v^2}\right)\, ,\quad \text{for}\,\, 1-2v^2 > 0\, ; \nonumber \\
    {} & = \textrm{arctan} \left(\frac{2i + i \kappa v + 2 \sqrt{-1 - \kappa v - v^4}}{\kappa v - 2v^2}\right)\, ,\quad \text{for}\,\, 1-2v^2 < 0\, , \label{eq:IYsol-3}
  \end{align}
  where $\ell$ is the length scale that appears in the solution from the formalism of Ref.~\cite{3D}, while $\ell_0$ is the length scale in the final expression. Apparently, $v = \frac{\sqrt{2}}{2}$ could be a singularity, but actually the results can be continued analytically to $v = \frac{\sqrt{2}}{2}$. Hence, it is not a real singularity. With this choice of parameters, we find:
  \begin{align}
    A_1 & = V_1 = -\frac{2 i u}{v \ell_0}\, ,\\
    H & = \frac{i}{v \ell_0}\, ,
  \end{align}
  where
  \begin{equation}
    u \equiv \sqrt{v^2 - 1}\, ,\quad \ell = \ell_0 (1 - 2v^2)\, .
  \end{equation}
  This is exactly the choice of the background auxiliary fields in Ref.~\cite{IY}.

\section{Localization of $\mathcal{N}=2$ Chern-Simons-Matter theory}
  \subsection{Review of the Theory}
  In this section, we briefly review the theory and the corresponding supersymmetry transformations and algebra constructed in Ref.~\cite{3D}, then in the next section, we will try to localize this theory on a squashed $S^3$ with $SU(2)\times U(1)$ isometry.

  As discussed in Ref.~\cite{3D}, the 3D $\mathcal{N}=2$ vector multiplet in the Wess-Zumino gauge transforms in the following way:
  \begin{align}
    \delta a_\mu & = -i (\zeta \gamma_\mu \widetilde{\lambda} + \widetilde{\zeta} \gamma_\mu \lambda)\, ,\nonumber\\
    \delta \sigma & = -\zeta \widetilde{\lambda} + \widetilde{\zeta} \lambda\, ,\nonumber\\
    \delta \lambda & = i \zeta (D + \sigma H) - \frac{i}{2} \varepsilon^{\mu\nu\rho} \gamma_\rho \zeta f_{\mu\nu} - \gamma^\mu\, \zeta (i \partial_\mu \sigma - V_\mu \sigma)\, ,\nonumber\\
    \delta \widetilde{\lambda} & = -i \widetilde{\zeta} (D+\sigma H) - \frac{i}{2} \varepsilon^{\mu\nu\rho} \gamma_\rho \widetilde{\zeta} f_{\mu\nu} + \gamma^\mu\, \widetilde{\zeta} (i\partial_\mu \sigma + V_\mu \sigma)\, ,\nonumber\\
    \delta D & = D_\mu (\zeta \gamma^\mu \widetilde{\lambda} - \widetilde{\zeta} \gamma^\mu \lambda) - iV_\mu (\zeta \gamma^\mu \widetilde{\lambda} + \widetilde{\zeta} \gamma^\mu \lambda) - H (\zeta \widetilde{\lambda} - \widetilde{\zeta} \lambda) + \zeta [\widetilde{\lambda},\, \sigma] - \widetilde{\zeta} [\lambda,\, \sigma]\, . \label{eq:fullSUSYgauge}
  \end{align}
  The transformations of the chiral multiplet and the anti-chiral multiplet are given by
  \begin{align}
    \delta \phi & = \sqrt{2} \zeta \psi\, ,\nonumber\\
    \delta \psi & = \sqrt{2} \zeta F - \sqrt{2} i (z - q \sigma - r H) \widetilde{\zeta} \phi - \sqrt{2} i \gamma^\mu \widetilde{\zeta} D_\mu \phi\, ,\nonumber\\
    \delta F & = \sqrt{2} i (z - q\sigma - (r-2) H) \widetilde{\zeta} \psi + 2i q \phi \widetilde{\zeta} \widetilde{\lambda} - \sqrt{2} i D_\mu (\widetilde{\zeta} \gamma^\mu \psi)\, ,\nonumber\\
    \delta \widetilde{\phi} & = -\sqrt{2} \widetilde{\zeta} \widetilde{\psi}\, ,\nonumber\\
    \delta \widetilde{\psi} & = \sqrt{2} \widetilde{\zeta} \widetilde{F} + \sqrt{2} i (z - q\sigma - rH) \zeta \widetilde{\phi} + \sqrt{2} i \gamma^\mu \zeta D_\mu \widetilde{\phi}\, ,\nonumber\\
    \delta \widetilde{F} & = \sqrt{2} i (z - q\sigma - (r-2) H) \zeta \widetilde{\psi} + 2 i q \widetilde{\phi} \zeta \lambda - \sqrt{2} i D_\mu (\zeta \gamma^\mu \widetilde{\psi})\, , \label{eq:fullSUSYmatter}
  \end{align}
  where $z$, $r$ and $q$ denote the central charge, the $R$-charge and the charge under the gauge group of the chiral multiplet repectively, and
  \begin{equation}
    D_\mu \equiv \nabla_\mu - ir (A_\mu - \frac{1}{2} V_\mu) - iz C_\mu - iq [a_\mu,\, \cdot]\, ,
  \end{equation}
  where $C_\mu$ satisfies
  \begin{equation}
    V^\mu = -i \varepsilon^{\mu\nu\rho} \partial_\nu C_\rho\, .
  \end{equation}
  The transformation parameters $\zeta$ and $\widetilde{\zeta}$ satisfy the two Killing spinor equations \eqref{eq:genKilling} with opposite $R$-charges respectively. Suppose that $\zeta$ and $\eta$ are two transformation parameters without tilde, and $\widetilde{\zeta}$ and $\widetilde{\eta}$ are two transformation parameters with tilde. It is checked in Ref.~\cite{3D} that the transformations with only parameters with tilde and only parameters without tilde satisfy the algebra:
  \begin{align}\label{eq:SUSYalg}
    \{\delta_\zeta,\, \delta_\eta\} \varphi & = 0\, , \nonumber\\
    \{\delta_{\widetilde{\zeta}},\, \delta_{\widetilde{\eta}}\} \varphi & = 0\, , \nonumber\\
    \{\delta_\zeta,\, \delta_{\widetilde{\zeta}}\} \varphi & = -2i \left(\mathcal{L}'_K \varphi + \zeta \widetilde{\zeta} (z - rH) \varphi \right)\, ,
  \end{align}
  where $\varphi$ denotes an arbitrary field, $K^\mu \equiv \zeta \gamma^\mu \widetilde{\zeta}$ and $\mathcal{L}'_K$ is a modified Lie derivative with the local $R$- and $z$-transformation
  \begin{equation}
    \mathcal{L}'_K \varphi \equiv \mathcal{L}_K \varphi - ir K^\mu (A_\mu - \frac{1}{2} V_\mu) \varphi - iz K^\mu C_\mu \varphi\, .
  \end{equation}

  Under these transformations, the following Lagrangians are supersymmetry invariant:
  \begin{enumerate}
    \item Fayet-Iliopoulos Term (for $U(1)$-factors of the gauge group):
          \begin{equation}\label{eq:Lag-1}
            \mathscr{L}_{FI} = \xi (D - a_\mu V^\mu - \sigma H)\, .
          \end{equation}
    \item Gauge-Gauge Chern-Simons Term:
          \begin{equation}\label{eq:Lag-2}
            \mathscr{L}_{gg} = \textrm{Tr} \left[\frac{k_{gg}}{4\pi} (i\varepsilon^{\mu\nu\rho} a_\mu \partial_\nu a_\rho - 2 D \sigma + 2i \widetilde{\lambda} \lambda) \right]\, .
          \end{equation}
    \item Gauge-$R$ Chern-Simons Term (for $U(1)$-factors of the gauge group):
          \begin{equation}\label{eq:Lag-3}
            \mathscr{L}_{gr} = \frac{k_{gr}}{2\pi} \left(i\varepsilon^{\mu\nu\rho} a_\mu \partial_\nu (A_\rho - \frac{1}{2} V_\rho) - DH + \frac{1}{4} \sigma (R - 2 V^\mu V_\mu - 2 H^2)\right)\, .
          \end{equation}
    \item Yang-Mills Term:
          \begin{align}\label{eq:Lag-4}
            \mathscr{L}_{YM} = & \textrm{Tr} \Bigg[ \frac{1}{4e^2} f^{\mu\nu} f_{\mu\nu} + \frac{1}{2e^2} \partial^\mu \sigma \partial_\mu \sigma - \frac{i}{e^2} \widetilde{\lambda} \gamma^\mu (D_\mu + \frac{i}{2} V_\mu) \lambda - \frac{i}{e^2} \widetilde{\lambda} [\sigma,\, \lambda]\nonumber\\
            {} & + \frac{i}{2e^2} \sigma \varepsilon^{\mu\nu\rho} V_\mu f_{\nu\rho} - \frac{1}{2e^2} V^\mu V_\mu \sigma^2 - \frac{1}{2e^2} (D + \sigma H)^2 + \frac{i}{2e^2} H \widetilde{\lambda} \lambda \Bigg]\, .
          \end{align}
     \item Matter Term:
           \begin{align}\label{eq:Lag-5}
             \mathscr{L}_{\textrm{mat}} = & \mathscr{D}^\mu \widetilde{\phi} \mathscr{D}_\mu \phi - i \widetilde{\psi} \gamma^\mu \mathscr{D}_\mu \psi - \widetilde{F} F + q (D + \sigma H) \widetilde{\phi} \phi - 2 (r - 1) H (z - q\sigma) \widetilde{\phi} \phi \nonumber\\
             {} & \left((z - q\sigma)^2 - \frac{r}{4} R + \frac{1}{2} (r - \frac{1}{2}) V^\mu V_\mu + r (r - \frac{1}{2}) H^2 \right) \widetilde{\phi} \phi \nonumber\\
             {} & \left(z - q\sigma (r - \frac{1}{2}) H \right) i \widetilde{\psi} \psi + \sqrt{2} i q (\widetilde{\phi} \lambda \psi + \phi \widetilde{\lambda} \widetilde{\psi})\, ,
           \end{align}
           where
           \begin{equation}
             \mathscr{D}_\mu \equiv \nabla_\mu - ir (A_\mu - \frac{1}{2} V_\mu) + i r_0 V_\mu - iz C_\mu - iq [a_\mu,\, \cdot]\, .
           \end{equation}
  \end{enumerate}
  In principle we could also add a superpotential term to the theory:
  \begin{equation}
    \int d^2 \theta\, W + \int d^2 \bar{\theta}\, \overline{W}\, ,
  \end{equation}
which is $\delta$-exact. The superpotential $W$ should be gauge invariant and have $R$-charge $2$, which imposes contraints on the fields and consequently affects the final result of the partition function. In this paper, for simplicity we do not consider a superpotential term.

  \subsection{Localization}
    To preserve the supersymmetry given by Eq.~\eqref{eq:fullSUSYgauge} and Eq.~\eqref{eq:fullSUSYmatter}, the following BPS equations should be satisfied:
    \begin{equation}\label{eq:LocalCond}
      Q \psi = 0\, ,\quad Q \widetilde{\psi} = 0\, ,\quad Q \lambda = 0\, ,\quad Q \widetilde{\lambda} = 0\, .
    \end{equation}
    In Appendix D, we show that these BPS equations lead to the classical solution
    \begin{equation}\label{eq:ClassSol}
      a_\mu = - \sigma C_\mu + a_\mu^{(0)}\, ,\quad \partial_\mu \sigma = 0\, ,\quad D = -\sigma H\, ,\quad \text{all other fields} = 0\, ,
    \end{equation}
    where $a_\mu^{(0)}$ is a flat connection, and $C_\mu$ appears in the new minimal supergravity as an Abelian gauge field, which satisfies $V^\mu = -i \varepsilon^{\mu\nu\rho} \partial_\nu C_\rho$ and still has the background gauge symmetry in this case:
    \begin{equation}\label{eq:bgdGaugeSymm}
      C_\mu \rightarrow C_\mu + \partial_\mu \Lambda^{(C)}\, .
    \end{equation}
    On the squashed $S^3$, $a_\mu^{(0)}$ can be set to $0$ by the gauge transformation. Moreover, since we have obtained $V_\mu$ before, we can also solve for $C_\mu$, but the solution is not unique due to the background gauge symmetry \eqref{eq:bgdGaugeSymm}. In the frame \eqref{eq:sqLeftinvFrame}, we can set $C_1 = 0$ by a background gauge transformation \eqref{eq:bgdGaugeSymm}. Hence, in this paper we have
    \begin{align}\label{eq:GaugeChoice}
      a_1 = 0\quad \Rightarrow \quad a_\mu V^\mu = 0 \quad \Rightarrow \quad & \varepsilon^{\mu\nu\rho} a_\mu \partial_\nu a_\rho \propto \varepsilon^{\mu\nu\rho} a_\mu \partial_\nu C_\rho \propto a_\mu V^\mu = 0\, ,\nonumber\\
      {} & \varepsilon^{\mu\nu\rho} a_\mu \partial_\nu (A_\rho - \frac{1}{2} V_\rho) \propto \varepsilon^{\mu\nu\rho} a_\mu \partial_\nu \eta_\rho \propto a_\mu (V^\mu - \kappa \eta^\mu) = 0\, ,
    \end{align}
    which will be relevant for later computations. These classical solutions give classical contributions to the partition function. We also need to consider the quantum fluctuation around these classical solutions, which will give 1-loop determinants to the partition function.

    The supersymmetry transformations introduced in the previous section is not nilpotent, and it is not obvious whether the supersymmetry invariant Lagrangians are also supersymmetry exact. It is more convenient to use a subset of the whole supersymmetry transformations to do the localization. In this paper we choose the subset to be the transformations without tilde, i.e. $\delta_\zeta$-transformations. For the matter sector the $\delta_\zeta$ transformations are
    \begin{align}
      Q \phi & \equiv \delta_\zeta \phi = \sqrt{2} \zeta \psi\, , \nonumber\\
      Q \psi & \equiv \delta_\zeta \psi = \sqrt{2} \zeta F\, , \nonumber\\
      Q F & \equiv \delta_\zeta F = 0\, , \nonumber\\
      Q \widetilde{\phi} & \equiv \delta_\zeta \widetilde{\phi} = 0\, , \nonumber\\
      Q \widetilde{\psi} & \equiv \delta_\zeta \widetilde{\psi} = \sqrt{2} i (z - q \sigma - r H) \zeta \widetilde{\phi} + \sqrt{2} i \gamma^\mu \zeta D_\mu \widetilde{\phi}\, , \nonumber\\
      Q \widetilde{F} & \equiv \delta_\zeta \widetilde{F} = \sqrt{2} i (z - q \sigma - (r-2) H) \zeta \widetilde{\psi} + 2 i q \zeta \lambda \widetilde{\phi} - \sqrt{2} i D_\mu (\zeta \gamma^\mu \widetilde{\psi})\, ,
    \end{align}
    while for the gauge sector the $\delta_\zeta$ transformations are
    \begin{align}
      Q a_\mu & \equiv \delta_\zeta a_\mu = -i \zeta \gamma_\mu \widetilde{\lambda}\, , \nonumber\\
      Q \sigma & \equiv \delta_\zeta \sigma = -\zeta \widetilde{\lambda}\, , \nonumber\\
      Q \lambda & \equiv \delta_\zeta \lambda = i \zeta (D + \sigma H) - \frac{i}{2} \varepsilon^{\mu\nu\rho} \gamma_\rho \zeta f_{\mu\nu} - \gamma^\mu \zeta (i \partial_\mu \sigma - V_\mu \sigma)\, , \nonumber\\
      Q \widetilde{\lambda} & \equiv \delta_\zeta \widetilde{\lambda} = 0\, , \nonumber\\
      Q D & \equiv \delta_\zeta D = \nabla_\mu (\zeta \gamma^\mu \widetilde{\lambda}) - i V_\mu (\zeta \gamma^\mu \widetilde{\lambda}) - H (\zeta \widetilde{\lambda}) + \zeta [\widetilde{\lambda},\, \sigma]\, .
    \end{align}

    From the supersymmetry algebra \eqref{eq:SUSYalg} we see that the $\delta_\zeta$-transformations are nilpotent. Then we can choose some $\delta_\zeta$-exact terms to localize the theory discussed in the previous section. For the matter sector we choose
    \begin{align}
      {} & \mathcal{V}_{\textrm{mat}} \equiv (Q \widetilde{\psi})^\dagger \widetilde{\psi} + (Q \psi)^\dagger \psi \\
      \Rightarrow\quad & Q \mathcal{V}_{\textrm{mat}} = (Q \widetilde{\psi})^\dagger (Q \widetilde{\psi}) + (Q \psi)^\dagger (Q \psi) - \widetilde{\psi} (Q (Q \widetilde{\psi})^\dagger) + Q (Q \psi)^\dagger \psi \\
      \Rightarrow\quad & Q(Q \mathcal{V}_{\textrm{mat}}) = 0\, .
    \end{align}
    For the gauge sector we choose
    \begin{align}
      {} & \mathcal{V}_g \equiv (Q \lambda)^\dagger \lambda + (Q \widetilde{\lambda})^\dagger \widetilde{\lambda} = (Q \lambda)^\dagger \lambda\\
      \Rightarrow\quad & Q \mathcal{V}_g = (Q \lambda)^\dagger (Q \lambda) + Q (Q \lambda)^\dagger \lambda\\
      \Rightarrow\quad & Q (Q \mathcal{V}_g) = 0 \, .
    \end{align}
    Precisely speaking, both $Q \mathcal{V}_{\textrm{mat}}$ and $Q \mathcal{V}_g$ will appear in the Lagrangian of the theory as $Q$-exact terms, and both of them contain the bosonic part and the fermionic part, i.e.,
    \begin{equation}
      Q \mathcal{V}_{\textrm{mat}} = (Q \mathcal{V}_{\textrm{mat}})_B + (Q \mathcal{V}_{\textrm{mat}})_F\, ,\quad Q \mathcal{V}_g = (Q \mathcal{V}_g)_B + (Q \mathcal{V}_g)_F\, ,
    \end{equation}
    where
    \begin{align}
      (Q \mathcal{V}_{\textrm{mat}})_B & \equiv (Q \widetilde{\psi})^\dagger (Q \widetilde{\psi}) + (Q \psi)^\dagger (Q \psi)\, , \nonumber\\
      (Q \mathcal{V}_{\textrm{mat}})_F & \equiv - \widetilde{\psi} (Q (Q \widetilde{\psi})^\dagger) + Q (Q \psi)^\dagger \psi\, , \nonumber\\
      (Q \mathcal{V}_g)_B & \equiv (Q \lambda)^\dagger (Q \lambda)\, , \nonumber\\
      (Q \mathcal{V}_g)_F & \equiv Q (Q \lambda)^\dagger \lambda\, .
    \end{align}

    For later convenience, we employ a trick similar to Ref.~\cite{MS-new} to rewrite the $Q$-transformations in the matter sector and $Q \mathcal{V}_{\textrm{mat}}$ in terms of a few operators. We will see that with the help of these operators the cancellation of the contributions from different modes to the partition function will be transparent.
    \begin{align}
      Q \phi & = - \sqrt{2} S_1^{c*} \psi\, , \nonumber\\
      Q \psi & = \sqrt{2} S_1 F\, , \nonumber\\
      Q F & = 0\, , \nonumber\\
      Q \widetilde{\phi} & = 0\, , \nonumber\\
      Q \widetilde{\psi} & = \sqrt{2} (S_2^c \phi)^\dagger\, , \nonumber\\
      Q \widetilde{F} & = \sqrt{2} (S_2^* \widetilde{\psi}^\dagger)^\dagger
    \end{align}
    where $\phi^\dagger \equiv \widetilde{\phi}$, and the operators $S_1$, $S_2$, $S_1^c$, $S_2^c$ and their corresponding adjoint operators $S_1^*$, $S_2^*$, $S_1^{c*}$, $S_2^{c*}$ are given by
    \begin{align}\label{eq:defOperators}
      S_1 \Phi & \equiv \Phi \zeta\, , \nonumber\\
      S_2 \Phi & \equiv -i \left[(z - q\sigma - (r-2) H) \Phi -  \slashed D \Phi \right] \zeta\, , \nonumber\\
      S_1^c \Phi & \equiv \Phi \zeta^\dagger\, , \nonumber\\
      S_2^c \Phi & \equiv i \zeta^\dagger \left[(\bar{z} - q \bar{\sigma} - r \bar{H}) + \slashed D \right] \Phi\, , \nonumber\\
      S_1^* \Psi & \equiv \zeta^\dagger \Psi\, , \nonumber\\
      S_2^* \Psi & \equiv i \zeta^\dagger \left[(\bar{z} - q\bar{\sigma} - (r - \frac{1}{2}) \bar{H}) - \frac{i}{2} V_\mu \gamma^\mu + \slashed D \right] \Psi\, , \nonumber\\
      S_1^{c*} \Psi & \equiv \zeta \Psi\, , \nonumber\\
      S_2^{c*} \Psi & \equiv - \left[i (z - q\sigma - (r-\frac{3}{2}) H) \zeta + \frac{1}{2} V_\mu \gamma^\mu \zeta + i \zeta \slashed D \right] \Psi
    \end{align}
    where $\Phi$ denotes an arbitrary bosonic field, while $\Psi$ denotes an arbitrary fermionic field. Direct computation shows that these operators satisfy the following orthogonality conditions:
    \begin{equation}\label{eq:orthogonalityCond}
      S_1^* S_1^c = 0 = S_1^{c*} S_1\, ,\quad S_2^* S_2^c = 0 = S_2^{c*} S_2\, .
    \end{equation}
    The derivation of the second relation is given in Appendix E. Moreover, the following relation turns out to be crucial later:
    \begin{equation}\label{eq:crucialRel}
      S_2 S_1^* + S_2^c S_1^{c*} = S_1 S_2^* + S_1^c S_2^{c*} - 2 i \textrm{Re}(z - q\sigma) e^{-2 \textrm{Im} \Theta} \Omega\, ,
    \end{equation}
    where $\Theta$ is the angle that appears in the rotation of the Killing spinor \eqref{eq:rotateKilling}. We prove the relation above in Appendix E. With these operators, one can show that
    \begin{align}
      \widetilde{\phi} \Delta_\phi \phi \equiv (Q \mathcal{V}_{\textrm{mat}})_B & = 2 \widetilde{\phi} S_2^{c*} S_2^c \phi\, , \nonumber\\
      \widetilde{\psi} \Delta_\psi \psi \equiv (Q \mathcal{V}_{\textrm{mat}})_F & = 2 \widetilde{\psi} (S_2 S_1^* + S_2^c S_1^{c*}) \psi\, .
    \end{align}
    Hence,
    \begin{equation}\label{eq:matterLaplacian}
      \Delta_\psi = 2 (S_2 S_1^* + S_2^c S_1^{c*})\, ,\quad \Delta_\phi = 2 S_2^{c*} S_2^c\, .
    \end{equation}

    For the gauge sector, instead of defining operators as in the matter sector, we can do direct computations, and the results are
    \begin{align}
      (Q \mathcal{V}_g)_B & = e^{-2 \textrm{Im} \Theta} \Omega \cdot \textrm{Tr} \left[-\frac{1}{2} f_{\mu\nu} f^{\mu\nu} - (D_\mu \sigma) (D^\mu \sigma) + (D + \sigma H)^2 \right]\, , \label{eq:gaugeLag-1}\\
      (Q \mathcal{V}_g)_F & = e^{-2 \textrm{Im} \Theta} \Omega \cdot \textrm{Tr} \left[2 i \widetilde{\lambda} \slashed D \lambda + 2 i [\widetilde{\lambda}, \sigma] \lambda - i H (\widetilde{\lambda} \lambda) - V_1 (\widetilde{\lambda} \lambda) - 2 V_\mu (\widetilde{\lambda} \gamma^\mu \lambda)\right]\, . \label{eq:gaugeLag-2}
    \end{align}

    \subsubsection{Classical Contribution}
    As discussed before, by inserting the classical solutions \eqref{eq:ClassSol} of the localization condition \eqref{eq:LocalCond} into the Lagrangians of the theory \eqref{eq:Lag-1}-\eqref{eq:Lag-5}, one obtains the classical contributions to the partition function. One can see immediately that $\mathscr{L}_{YM}$ and $\mathscr{L}_{\textrm{mat}}$ do not have classical contributions to the partition function. Due to Eq.~\eqref{eq:GaugeChoice}, the classical contributions from other Lagrangians also simplify to be
    \begin{align}
      \textrm{exp} \left(i \int d^3 x \sqrt{g} \mathscr{L}_{FI} \right) & = \textrm{exp} \left(i \xi \int d^3 x \sqrt{g}\, \textrm{Tr}[D - \sigma H] \right) \nonumber\\
      {} & = \textrm{exp} \left( -\frac{4i \pi^2 \xi \ell^3}{v} H\, \textrm{Tr} (\sigma) \right)\, ,\label{eq:Zclass-1}\\
      \textrm{exp} \left(i \int d^3 x \sqrt{g} \mathscr{L}_{gg} \right) & = \textrm{exp} \left(\frac{i k_{gg}}{4\pi} \int d^3 x \sqrt{g}\, \textrm{Tr}[-2 D \sigma] \right) \nonumber\\
      {} & = \textrm{exp} \left( \frac{i\pi k_{gg} \ell^3}{v} H\, \textrm{Tr} (\sigma^2) \right)\, ,\label{eq:Zclass-2}\\
      \textrm{exp} \left(i \int d^3 x \sqrt{g} \mathscr{L}_{gr} \right) & = \textrm{exp} \left(\frac{i k_{gr}}{2\pi} \int d^3 x \sqrt{g}\, \textrm{Tr} \left[-D H + \frac{1}{4} \sigma (R - 2 V^\mu V_\mu - 2 H^2) \right] \right) \nonumber\\
      {} & = \textrm{exp} \left( \frac{i \pi k_{gr} \ell^3}{2v} (H^2 + \frac{1}{2}R - V_\mu V^\mu)\, \textrm{Tr} (\sigma) \right)\, ,\label{eq:Zclass-3}
    \end{align}
    where $H$ and $V_\mu$ are in general the auxiliary fields after shifting which are given by Eq.~\eqref{eq:genAux}, and $R = \frac{8}{\ell^2} - \frac{2}{\ell^2 v^2}$ is the Ricci scalar of the squashed $S^3$ considered in this paper.

    \subsubsection{1-Loop Determinant for Matter Sector}
    The key step of localization is to calculate the 1-loop determinants for the gauge sector and the matter sector. There are a few methods available to do this step:
    \begin{itemize}
      \item Use the index theorem;
      \item Expand the Laplacians into spherical harmonics;
      \item Consider the modes that are not paired and consequently have net contributions to the partition function.
    \end{itemize}
    All three methods have been used in many papers. In our case, it is more convenient to use the third one, which originated in Ref.~\cite{Jap-2} and has been done in a more systematic way in Ref.~\cite{MS-new}. In this and next subsection, I will follow closely the method in Ref.~\cite{MS-new} and apply it to our case of interest.

    The basic idea is to first find out how the modes are paired, since paired modes cancel out exactly and do not have contributions to the partition function. If in a pair the fermionic partner is missing, which is called missing spinor, then the bosonic partner has a net contribution to the denominator of the 1-loop determinant in the partition function. If in a pair the bosonic partner is missing, then the fermionic partner is called unpaired spinor, and has a net contribution to the numerator of the 1-loop determinant in the partition function.

    Starting from Eq.~\eqref{eq:matterLaplacian}, we assume that
    \begin{equation}
      \frac{1}{2} \Delta_\phi \Phi = S_2^{c*} S_2^c \Phi = \mu \Phi\, ,
    \end{equation}
    and define
    \begin{equation}
      \Psi_1 \equiv S_1^c \Phi\, ,\quad \Psi_2 \equiv S_2^c \Phi\, ,
    \end{equation}
    then using Eq.~\eqref{eq:matterLaplacian} and Eq.~\eqref{eq:crucialRel} we obtain
    \begin{align}
      \frac{1}{2} \Delta_\psi \Psi_1 & = (S_2 S_1^* + S_2^c S_1^{c*}) S_1^c \Phi \nonumber\\
      {} & = S_2^c S_1^{c*} S_1^c \Phi \nonumber\\
      {} & = -e^{-2 \textrm{Im} \Theta} \Omega \Psi_2\, , \\
      \frac{1}{2} \Delta_\psi \Psi_2 & = \left[S_1 S_2^* + S_1^c S_2^{c*} - 2 i \textrm{Re} (z - q\sigma) e^{-2 \textrm{Im} \Theta} \Omega \right] S_2^c \Phi \nonumber\\
      {} & = S_1^c \mu \Phi - 2i \textrm{Re} (z - q\sigma) e^{-2 \textrm{Im} \Theta} \Omega S_2^c \Phi \nonumber\\
      {} & = \mu \Psi_1 - 2 i A e^{-2 \textrm{Im} \Theta} \Psi_2\, ,
    \end{align}
    where $A \equiv \textrm{Re} (z - q\sigma)\cdot \Omega$. I.e.,
    \begin{equation}\label{eq:spinorModes}
      \Delta_\psi \left(\begin{array}{c} \Psi_1 \\ \Psi_2 \end{array}\right) = \left(\begin{array}{cc}
                                                                                       0 & -e^{-2 \textrm{Im} \Theta} \Omega\\
                                                                                       \mu & -2i A e^{-2 \textrm{Im}\Theta}
                                                                                     \end{array}\right) \cdot \left(\begin{array}{c} \Psi_1 \\ \Psi_2 \end{array}\right)\, .
    \end{equation}
    The eigenvalues of $\Delta_\psi$ in this subspace are
    \begin{equation}
      \lambda_{1,2} = e^{-2 \textrm{Im}\Theta} \left[-i A \pm \sqrt{-A^2 - \widetilde{\mu}} \right]\, ,
    \end{equation}
    where $\widetilde{\mu} \equiv \mu \Omega e^{2 \textrm{Im}\Theta}$. Suppose that $\widetilde{\mu} = - M^2 + 2i A M$, then
    \begin{equation}
      \lambda_{1,2} = -e^{-2 \textrm{Im}\Theta} M\, ,\quad e^{-2 \textrm{Im}\Theta} (M - 2iA)\, .
    \end{equation}
    In other words, if there exists a bosonic mode $\Phi$ satisfying
    \begin{align}
      \Delta_\phi \Phi = \mu \Phi & = \frac{1}{\Omega} e^{-2 \textrm{Im} \Theta} \widetilde{\mu} \Phi \nonumber\\
      {} & = \frac{1}{\Omega} e^{-2 \textrm{Im} \Theta} (-M) (M - 2iA) \Phi\, ,
    \end{align}
    there are corresponding fermionic modes $\Psi_1$ and $\Psi_2$, which span a subspace, and $\Delta_\psi$ has eigenvalues $\lambda_{1,2}$ with
    \begin{equation}
      \lambda_1 \cdot \lambda_2 = e^{-4 \textrm{Im}\Theta} (-M) (M - 2iA)\, .
    \end{equation}
    To make the modes paired, we can rescale the bosonic mode $\Phi$ appropriately, or equivalently define
    \begin{equation}
      \hat{\Delta}_\phi \equiv e^{-2 \textrm{Im} \Theta} \Omega \Delta_\phi\, ,
    \end{equation}
    then
    \begin{equation}
      \hat{\Delta}_\phi \Phi = e^{-4 \textrm{Im}\Theta} (-M) (M - 2iA) \Phi\, .
    \end{equation}
    Therefore, if there are no missing spinors or unpaired spinors, the bosonic modes and the fermionic modes cancel out exactly.

    Conversely, if there exists a fermionic mode $\Psi$ satisfying
    \begin{equation}
      \Delta_\psi \Psi = -M e^{-2 \textrm{Im} \Theta} \Psi\, ,
    \end{equation}
    then using Eq.~\eqref{eq:crucialRel} we can rewrite the condition above as
    \begin{equation}
      S_1 S_2^* \Psi + S_1^c S_2^{c*} \Psi = (2iA - M) e^{-2 \textrm{Im} \Theta} \Psi\, ,
    \end{equation}
    where recall $A \equiv \textrm{Re}(z - q\sigma)\cdot \Omega$. Acting $S_1^{c*}$ from the left and using the orthogonality condition \eqref{eq:orthogonalityCond}, we obtain
    \begin{equation}\label{eq:tempRel}
      S_2^{c*} \Psi = -\frac{1}{\Omega} (2iA - M) \Phi\, .
    \end{equation}
    Acting $S_2^{c*}$ from left on the equation
    \begin{equation}
      \Delta_\psi \Psi = (S_2 S_1^* + S_2^c S_1^{c*}) \Psi = -M e^{-2 \textrm{Im} \Theta} \Psi\, ,
    \end{equation}
    using the relation \eqref{eq:tempRel} we just obtained, we find
    \begin{equation}
      \Delta_\phi \Phi = S_2^{c*} S_2^c (S_1^{c*} \Psi) = -\frac{1}{\Omega} e^{-2 \textrm{Im} \Theta} M (M - 2iA) \Phi\, .
    \end{equation}
    I.e., for a fermionic mode $\Psi$, the corresponding bosonic mode can be constructed as $\Phi = S_1^{c*} \Psi$.

    With the preparation above, we can consider the unpaired spinors and the missing spinors. For the unpaired spinor, there is no corresponding bosonic partner, i.e.,
    \begin{displaymath}
      \Phi = S_1^{c*} \Psi = 0\, . 
    \end{displaymath}
    Based on the orthogonality condition \eqref{eq:orthogonalityCond} there should be
    \begin{equation}
      \Psi = S_1 F\, .
    \end{equation}
    Then
    \begin{align}\label{eq:Mpsi}
      {} & \Delta_\psi \Psi = S_2 S_1^* \Psi = M_\psi \Psi \nonumber\\
      \Rightarrow \quad & S_2 S_1^* S_1 F = M_\psi S_1 F \nonumber\\
      \Rightarrow \quad & e^{-2 \textrm{Im} \Theta} \Omega S_2 F = M_\psi S_1 F \nonumber\\
      \Rightarrow \quad & e^{-2 \textrm{Im} \Theta} \Omega S_1^* S_2 F = M_\psi S_1^* S_1 F \nonumber\\
      \Rightarrow \quad & M_\psi = i e^{-2 \textrm{Im} \Theta} \Omega [D_1 - z + q \sigma + (r-2) H]\, .
    \end{align}
    For the missing spinor, $\Psi_2 \propto \Psi_1$. Suppose that
    \begin{equation}
      \Psi_2 = a \Psi_1\, ,
    \end{equation}
    where $a$ is a constant. Then from Eq.~\eqref{eq:spinorModes} we know that
    \begin{align}
      \Delta_\psi \Psi_2 & = a \Delta_\psi \Psi_1 = (\mu - 2iA e^{-2 \textrm{Im}\Theta} a) \Psi_1 \\
      \Delta_\psi \Psi_1 & = -e^{-2 \textrm{Im} \Theta} \Omega a \Psi_1
    \end{align}
    \begin{align}
      \Rightarrow \quad & \frac{\Delta_\psi \Psi_1}{\Psi_1} = \frac{\mu - 2iA e^{-2 \textrm{Im} \Theta} a}{a} = -e^{-2 \textrm{Im} \Theta} \Omega a \nonumber\\
      \Rightarrow \quad & \mu = 2iA e^{-2 \textrm{Im} \Theta} a - e^{-2 \textrm{Im} \Theta} \Omega a^2 = \frac{1}{\Omega} e^{-2 \textrm{Im} \Theta} (-M^2 + 2iAM) \nonumber\\
      \Rightarrow \quad & a = \frac{M}{\Omega}\, .
    \end{align}
    Hence,
    \begin{equation}
      \Psi_2 = \frac{M}{\Omega} \Psi_1\, ,
    \end{equation}
    i.e.,
    \begin{equation}
      S_2^c \Phi = \frac{M}{\Omega} S_1^c \Phi\, .
    \end{equation}
    Acting $S_1^{c*}$ from left, we obtain
    \begin{align}\label{eq:Mphi}
      {} & S_1^{c*} S_2^c \Phi = \frac{M_\phi}{\Omega} S_1^{c*} S_1^c \Phi \nonumber\\
      \Rightarrow \quad & -i (\bar{z} - q \bar{\sigma} - r \bar{H}) (\zeta^\dagger \zeta) \phi - i (\zeta^\dagger \gamma^\mu \zeta) D_\mu \phi = - \frac{M_\phi}{\Omega} (\zeta^\dagger \zeta) \phi \nonumber\\
      \Rightarrow \quad & M_\phi = \Omega [iD_1 + i (\bar{z} - q \bar{\sigma} - r \bar{H})]\, .
    \end{align}

    To proceed, we have to figure out the eigenvalues of the operator $D_1$ in $M_\psi$ \eqref{eq:Mpsi} and $M_\phi$ \eqref{eq:Mphi}. Similar to Ref.~\cite{IY}, we use
    \begin{displaymath}
      |s,\, s_z\rangle \quad \textrm{with}\,\, s_z = -s,\, -s+1,\, \cdots,\, s-1,\, s
    \end{displaymath}
    as the spin basis, which transforms in the $(0,\, s)$ representation of $SU(2)_L\times SU(2)_R$ and
    \begin{displaymath}
      |j,\, m',\, m\rangle = Y_{m',\, m}^j
    \end{displaymath}
    as the orbital basis, which transforms in the $(j,\, j)$ representation of $SU(2)_L\times SU(2)_R$, where $j$ is the azimuthal quantum number, while $m'$ and $m$ are the magnetic quantum numbers for $SU(2)_L$ and $SU(2)_R$ respectively, and they take values in the following ranges:
    \begin{align}
      j & = 0,\, \frac{1}{2},\, 1,\, \cdots\, ; \nonumber\\
      m' & = -j,\, -j+1,\, \cdots ,\, j-1,\, j\, ; \nonumber\\
      m & = -j,\, -j+1,\, \cdots ,\, j-1,\, j\, .
    \end{align}
    For a group element $g \in SU(2)$, a field $\Phi(g)$ can be expanded as
    \begin{equation}
      \Phi(g) = \sum_{j,\, m',\, m,\, s_z} \Phi_{m',\, m,\, s_z}^j |j,\, m',\, m\rangle \otimes |s,\, s_z\rangle\, .
    \end{equation}

    The covariant derivative on $S^3$ can be written as
    \begin{equation}\label{eq:covDerS3-left}
      \nabla^{(0)} = \mu^1 (2 L^1 - S^1) + \mu^2 (2 L^2 - S^2) + \mu^3 (2 L^3 - S^3)\, ,
    \end{equation}
    where $L_m$ denote the orbital angular momentum operators on $SU(2)_R$, while $S_m$ are the spin operators. Similarly, it can also be written in the right-invariant frame as
    \begin{equation}\label{eq:covDerS3-right}
      \nabla^{(0)} = \widetilde{\mu}\,^1 (2 L'^1 + S^1) + \widetilde{\mu}\,^2 (2 L'^2 + S^2) + \widetilde{\mu}\,^3 (2 L'^3 + S^3)\, ,
    \end{equation}
    where $L'_m$ denote the orbital angular momentum operators on $SU(2)_L$, while $S_m$ are still the spin operators. More generally, we can write the covariant derivative $D^{(0)}$ as a combination of the expressions \eqref{eq:covDerS3-left} and \eqref{eq:covDerS3-right}:
    \begin{align}
      \nabla^{(0)} = & \quad a \left[\mu^1 (2 L^1 - S^1) + \mu^2 (2 L^2 - S^2) + \mu^3 (2 L^3 - S^3)\right] \nonumber\\
      {} & \quad + (1-a) \left[\widetilde{\mu}\,^1 (2 L'^1 + S^1) + \widetilde{\mu}\,^2 (2 L'^2 + S^2) + \widetilde{\mu}\,^3 (2 L'^3 + S^3) \right]\, ,
    \end{align}
    where $a$ is an arbitrary constant. For the squashed $S^3$ with $SU(2)\times U(1)$ isometry given by the metric \eqref{eq:squashedMetricInForms}, the covariant derivative also has different expressions as follows:
    \begin{align}
      \nabla = & \quad \mu^1 \left(2 L^1 - (2 - \frac{1}{v^2}) S^1\right) + \mu^2 (2 L^2 - \frac{1}{v} S^2) + \mu^3 (2 L^3 - \frac{1}{v} S^3) \label{eq:covDerSqS3-left}\\
      = & \quad \widetilde{\mu}\,^1 \left(2 L'^1 + (2 - \frac{1}{v^2}) S^1\right) + \widetilde{\mu}\,^2 (2 L'^2 + \frac{1}{v} S^2) + \widetilde{\mu}\,^3 (2 L'^3 + \frac{1}{v} S^3) \label{eq:covDerSqS3-right}\\
      = & \quad a \left[\mu^1 \left(2 L^1 - (2 - \frac{1}{v^2}) S^1\right) + \mu^2 (2 L^2 - S^2) + \mu^3 (2 L^3 - S^3)\right] \nonumber\\
      {} & \quad + (1-a) \left[\widetilde{\mu}\,^1 \left(2 L'^1 + (2 - \frac{1}{v^2}) S^1\right) + \widetilde{\mu}\,^2 (2 L'^2 + S^2) + \widetilde{\mu}\,^3 (2 L'^3 + S^3) \right]\, , \label{eq:covDerSqS3-mix}
    \end{align}
    where $a$ again can be an arbitrary constant.

    Since the squashed $S^3$ that we consider in this paper has $SU(2)_L \times U(1)_R$ isometry, $L'_m L'_m$, $L'_1$ and $L_1 + S_1$ should have well-defined eigenvalues as follows:
    \begin{equation}
      L'_m L'_m = -j(j+1)\, ,\quad L'_1 = im'\, ,\quad L_1 + S_1 = im\, .
    \end{equation}
    Knowing this, we can return to the discussion of the eigenvalues of the Laplacians $\Delta_\psi$ and $\Delta_\phi$ \eqref{eq:Mpsi} \eqref{eq:Mphi}. Remember that both expressions are derived from some scalar modes, hence the spin $s = 0$, i.e., $S_1$ has vanishing eigenvalues. Then the covariant derivative without the gauge connection, i.e. $\nabla_1$, in both expressions has the form
    \begin{equation}\label{eq:nablaTrivialCase}
      \nabla_1 = \frac{v}{\ell}\cdot 2 L_1 = \frac{v}{\ell}\cdot 2im
    \end{equation}
    on the states $|j,\, m',\, m\rangle$ with $-j \leqslant m',\, m \leqslant j$ and $j = 0,\, \frac{1}{2},\, 1,\, \cdots$ of $SU(2)$. Hence, the eigenvalues of the Laplacians $\Delta_\psi$ and $\Delta_\phi$ \eqref{eq:Mpsi} \eqref{eq:Mphi} can be expressed as
    \begin{align}
      M_\psi & = i e^{-2 \textrm{Im} \Theta} \Omega \left[\nabla_1 - i(r-2) (A_1 - \frac{1}{2} V_1) - i(z - q\sigma) C_1 - (z - q\sigma) + (r-2) H \right] \nonumber\\
      {} & = i e^{-2 \textrm{Im} \Theta} \Omega \left[\frac{2iv}{\ell} m - i (r-2) (A_1 - \frac{1}{2} V_1 + iH) - (z - q\sigma) \right]\, ,\\
      M_\phi & = i \Omega \left[\nabla_1 - ir (A_1 - \frac{1}{2} V_1) - i(z - q\sigma) C_1 + (\bar{z} - q\bar{\sigma}) + rH \right] \nonumber\\
      {} & = i \Omega \left[\frac{2iv}{\ell} m - ir (A_1 - \frac{1}{2} V_1 + iH) + (\bar{z} - q \bar{\sigma}) \right]\, ,
    \end{align}
    where we have used the background gauge symmetry \eqref{eq:bgdGaugeSymm} to set $C_1 = 0$, as mentioned before. In general, $z$ and $\sigma$ can be complex, which is crucial in some cases \cite{3D}, in this paper for simplicity we assume that
    \begin{equation}
      \bar{\sigma} = -\sigma\, ,\quad \bar{z} = -z\, .
    \end{equation}
    As a check, let us consider a few previously studied cases. For round $S^3$, there are
    \begin{displaymath}
      v=1\, ,\quad A_1 = V_1 = 0\, ,\quad H = -\frac{i}{\ell}\, ,\quad z = 0\, ,
    \end{displaymath}
    then the 1-loop determinant for the matter sector is
    \begin{align}
      Z_{\textrm{mat}}^{1-\textrm{loop}} & = \prod_{\rho \in R} \prod_{j=0}^\infty \left( \prod_{m=-j}^j \frac{\frac{2im}{\ell} - i(r-2)\frac{1}{\ell} + q\rho(\sigma)}{\frac{2im}{\ell} - ir\frac{1}{\ell} + q \rho(\sigma)} \right)^{2j+1} \nonumber\\
      {} & = \prod_{\rho \in R} \prod_{n=0}^\infty \left(- \frac{\frac{n+1-r}{\ell} + i \rho(\sigma)}{\frac{n-1+r}{\ell} - i\rho(\sigma)} \right)^n\, .
    \end{align}
    where $\rho$ denotes the weights in the representation $R$, and we have set $q = -1$ and $n \equiv 2j+1$. This result is precisely the one obtained in Ref.~\cite{Kapustin-1}. Similarly, if we choose
    \begin{displaymath}
      A_1 = \frac{v}{\ell} - \frac{1}{v\ell}\, ,\quad V_1 = 0\, ,\quad H = -\frac{i}{v\ell}\, ,\quad  z=0\, ,
    \end{displaymath}
    the 1-loop determinant for the matter sector becomes
    \begin{align}\label{eq:matterPartHHL}
      Z_{\textrm{mat}}^{1-\textrm{loop}} & = \prod_{\rho \in R} \prod_{j=0}^\infty \left( \prod_{m=-j}^j \frac{\frac{2ivm}{\ell} - i(r-2) \frac{v}{\ell} + q \rho(\sigma)}{\frac{2ivm}{\ell} - ir \frac{v}{\ell} + q \rho(\sigma)} \right)^{2j+1} \nonumber\\
      {} & = \prod_{\rho \in R} \prod_{j=0}^\infty \left(- \frac{\frac{2j+2-r}{\tilde{\ell}} + i\rho(\sigma)}{\frac{2j+r}{\tilde{\ell}} - i\rho(\sigma)} \right)^{2j+1}\, ,
    \end{align}
    where $\tilde{\ell} \equiv \frac{\ell}{v}$, and this result is the same as the one with $SU(2)\times U(1)$ isometry in Ref.~\cite{Jap-2}. An immediate generalization is to shift the auxiliary fields by $\sim \kappa$ without rotating the Killing spinors, i.e., for
    \begin{displaymath}
      A_1 = \frac{v}{\ell} + \frac{2}{v\ell} + \frac{3\kappa}{2\ell}\, ,\quad V_1 = \frac{2}{v\ell} + \frac{\kappa}{\ell}\, ,\quad H = \frac{i}{v\ell} + i \frac{\kappa}{\ell}\, ,\quad z=0\, ,
    \end{displaymath}
    the partition function remains the same as Eq.~\eqref{eq:matterPartHHL}, i.e., the one in Ref.~\cite{Jap-2}:
    \begin{equation}
      Z_{\textrm{mat}}^{1-\textrm{loop}} = \prod_{\rho \in R} \prod_{j=0}^\infty \left(- \frac{\frac{2j+2-r}{\tilde{\ell}} + i\rho(\sigma)}{\frac{2j+r}{\tilde{\ell}} - i\rho(\sigma)} \right)^{2j+1}\, .
    \end{equation}
    At this point, we may conclude that the shift $\kappa$ in the auxiliary fields does not affect the 1-loop determinant of the matter sector, and to obtain a nontrivial result like the one in Ref.~\cite{IY} one has to consider the rotation of the Killing spinors \eqref{eq:genAux}, i.e., $\Theta \neq 0$. All the examples discussed above can be thought of to be $\Theta = 0$.

    For the cases with $\Theta \neq 0$, e.g. the case discussed in Ref.~\cite{IY}, the main difference is that
    \begin{equation}
      A_1 - \frac{1}{2} V_1 + iH \neq \frac{v}{\ell}\, .
    \end{equation}
    For Ref.~\cite{IY} there is
    \begin{equation}
      A_1 - \frac{1}{2} V_1 + iH = - \frac{1}{\ell} \left(\frac{\sqrt{1-v^2}}{v} + \frac{1}{v} \right) = -\frac{1+iu}{v\ell}\, .
    \end{equation}
    This change will affect the expressions of $M_\psi$ \eqref{eq:Mpsi} and $M_\phi$ \eqref{eq:Mphi}. We can think of this effect as to use the background gauge fields to twist the connections in the covariant derivatives, i.e., we want to absorb the background gauge fields into the covariant derivatives. For $\Theta = 0$, there is always $A_1 - \frac{1}{2} V_1 + iH = \frac{v}{\ell}$. From Eq.~\eqref{eq:nablaTrivialCase} we can see that the background fields can be thought of to only twist the $SU(2)_L$ part of the connection, without affecting the $SU(2)_R$ part of the connection. For $\Theta \neq 0$, $A_1 - \frac{1}{2} V_1 + iH$ is in general not equal to $\frac{v}{\ell}$, hence cannot be absorbed only in the $SU(2)_L$ part of the connection, i.e., it has to twist also the $SU(2)_R$ part. We can use the expression \eqref{eq:covDerSqS3-mix} and figure out the coefficients in the linear combination. The guiding principle is that we still want to require that the background gauge fields can be absorbed only in the $SU(2)_L$ part of the connection. Hence, from Eq.~\eqref{eq:covDerSqS3-mix} and Eq.~\eqref{eq:nablaTrivialCase} we see that on the states $|j,\, m',\, m\rangle$ with the spin $s=0$:
    \begin{displaymath}
      \nabla_1 - 2i (A_1 - \frac{1}{2} V_1 + iH) = \left[a \frac{v}{\ell} 2 L^1 + (1-a) \frac{v}{\ell} 2 L'^1 \right] - 2i \frac{v}{\ell_1} = \left[a \frac{v}{\ell} 2 im + (1-a) \frac{v}{\ell} 2 im' \right] - 2i \frac{v}{\ell_1}\, ,
    \end{displaymath}
    where
    \begin{equation}
      \frac{v}{\ell_1} \equiv \pm (A_1 - \frac{1}{2} V_1 + iH)\, .
    \end{equation}
    Since $-j \leqslant m',\, m \leqslant j$, the choice of the sign does not change the final result. To combine $2i\frac{v}{\ell_1}$ with the term containing $L^1$ implies that
    \begin{displaymath}
      a \frac{v}{\ell} = \frac{v}{\ell_1}\quad \Rightarrow \quad a = \frac{\ell}{\ell_1}\, .
    \end{displaymath}
    Then
    \begin{equation}\label{eq:nablaAfterTwist}
      \nabla_1 - 2i (A_1 - \frac{1}{2} V_1 + iH) = \left[\frac{v}{\ell_1} 2 L^1 + (\frac{v}{\ell} - \frac{v}{\ell_1}) 2 L'^1\right] - 2i \frac{v}{\ell_1} = \left[\frac{v}{\ell_1} 2 im + (\frac{v}{\ell} - \frac{v}{\ell_1}) 2 im'\right] - 2i \frac{v}{\ell_1}\, .
    \end{equation}
    For the case discussed in Ref.~\cite{IY} there is
    \begin{align}
      {} & \frac{v}{\ell_1} = \frac{1+iu}{v\ell} = \frac{v}{(1-iu)\ell}\\
      \Rightarrow\quad & \frac{v}{\ell} - \frac{v}{\ell_1} = \frac{v}{\ell} - \frac{v}{(1-iu)\ell} = \frac{-iu}{1-iu} \frac{v}{\ell} = -iu \frac{v}{\ell_1}\, .
    \end{align}
    Then the 1-loop determinant for the matter sector has the general expression
    \begin{align}
      Z_{\textrm{mat}}^{1-\textrm{loop}} & = \prod_{\rho \in R} \, \prod_j \prod_{-j \leqslant m,\, m' \leqslant j} e^{-2 \textrm{Im} \Theta}\, \frac{\frac{2iv}{\ell_1} (m-1) + 2i \left(\frac{v}{\ell} - \frac{v}{\ell_1} \right) m' + ir \frac{v}{\ell_1} - (z - q\rho(\sigma))}{\frac{2iv}{\ell_1} m + 2i \left(\frac{v}{\ell} - \frac{v}{\ell_1} \right) m' + ir \frac{v}{\ell_1} + (\bar{z} - q\rho(\bar{\sigma}_0))} \nonumber\\
      {} & = \prod_{\rho \in R} \, \prod_j \prod_{-j \leqslant m,\, m' \leqslant j} e^{-2 \textrm{Im} \Theta}\, \frac{2i (m-1) + 2u m' + ir + \frac{q\ell \rho(\sigma)}{b}}{2im + 2u m' + ir + \frac{q\ell \rho(\sigma)}{b}} \nonumber\\
      {} & = \prod_{\rho \in R} \, \prod_j \prod_{-j \leqslant m' \leqslant j} e^{-2 \textrm{Im} \Theta}\, \frac{2i (-j-1) + 2u m' + ir + \frac{q\ell \rho(\sigma)}{b}}{2ij + 2u m' + ir + \frac{q\ell \rho(\sigma)}{b}} \nonumber\\
      {} & = \prod_{\rho \in R} \, \prod_j \prod_{-j \leqslant m' \leqslant j} e^{-2 \textrm{Im} \Theta}\,  \left( -\frac{2j + 2 + 2iu m' - r + \frac{iq\ell \rho(\sigma)}{b}}{2j - 2iu m' + r - \frac{iq\ell \rho(\sigma)}{b}}\right)\, ,
    \end{align}
    where $b \equiv \frac{1+iu}{v}$, $\rho$ again denotes the weights in the representation $R$, and we have assumed
    \begin{displaymath}
      z = 0\, ,\quad \bar{\sigma} = -\sigma\, .
    \end{displaymath}
    If we identify $r$ and $\frac{\ell}{b}$ with $\Delta$ and $r$ in Ref.~\cite{IY} respectively, and let $q=1$, then up to some constant this result is the same as the one in Ref.~\cite{IY}.

    For the most general auxiliary fields \eqref{eq:genAux} on a squashed $S^3$ with $SU(2)\times U(1)$ isometry, the 1-loop determinant for the matter sector is
    \begin{align}
      Z_{\textrm{mat}}^{1-\textrm{loop}} & = \prod_{\rho \in R} \, \prod_j \prod_{-j \leqslant m,\, m' \leqslant j} e^{-2 \textrm{Im} \Theta}\, \frac{2i (m-1) \frac{v}{\ell_1} + 2i m' \left(\frac{v}{\ell} - \frac{v}{\ell_1} \right) + ir \frac{v}{\ell_1} - (z - q \rho(\sigma))}{2im \frac{v}{\ell_1} + 2im' \left(\frac{v}{\ell} - \frac{v}{\ell_1} \right) + ir \frac{v}{\ell_1} + (\bar{z} - q \rho(\bar{\sigma}_0))} \nonumber\\
      {} & = \prod_{\rho \in R} \, \prod_j \prod_{-j \leqslant m' \leqslant j} e^{-2 \textrm{Im} \Theta}\, \left(- \frac{2j + 2 - 2m'\, \frac{\frac{v}{\ell} - \frac{v}{\ell_1}}{\frac{v}{\ell_1}} - r - i\, \frac{z - q \rho(\sigma)}{\frac{v}{\ell_1}}}{2j + 2m'\, \frac{\frac{v}{\ell} - \frac{v}{\ell_1}}{\frac{v}{\ell_1}} + r + i\, \frac{z - q \rho(\sigma)}{\frac{v}{\ell_1}}} \right) \nonumber\\
      {} & = \prod_{\rho \in R} \, \prod_{p,\, q = 0}^\infty e^{-2 \textrm{Im} \Theta}\, \left(- \frac{p + q + 2 - (p-q) W - r - i\, \frac{z - q \rho(\sigma)}{\frac{v}{\ell_1}}}{p + q + (p - q) W + r + i\, \frac{z - q \rho(\sigma)}{\frac{v}{\ell_1}}} \right) \nonumber\\
      {} & = \prod_{\rho \in R} \, \prod_{p,\, q = 0}^\infty e^{-2 \textrm{Im} \Theta}\, \left(- \frac{(1-W) p + (1+W) q + 1 - i \left(\frac{z - q\rho(\sigma)}{\frac{v}{\ell_1}} - ir + i \right)}{(1+W) p + (1-W) q + 1 + i \left(\frac{z - q\rho(\sigma)}{\frac{v}{\ell_1}} - ir + i \right)} \right) \nonumber\\
      {} & = \prod_{\rho \in R} \, \prod_{p,\, q = 0}^\infty e^{-2 \textrm{Im} \Theta}\, \left(- \frac{b p + b^{-1} q + \frac{b+b^{-1}}{2} - \frac{i(b+b^{-1})}{2} \left(\frac{z - q\rho(\sigma)}{\frac{v}{\ell_1}} - ir + i \right)}{b^{-1} p + bq + \frac{b+b^{-1}}{2} + \frac{i(b+b^{-1})}{2} \left(\frac{z - q\rho(\sigma)}{\frac{v}{\ell_1}} - ir + i \right)} \right)\, ,
    \end{align}
    where
    \begin{equation}\label{eq:def-jm}
      j = \frac{p + q}{2}\, ,\quad m' = \frac{p - q}{2}\, ,
    \end{equation}
    \begin{equation}\label{eq:def-Wb}
      W \equiv \frac{\frac{v}{\ell} - \frac{v}{\ell_1}}{\frac{v}{\ell_1}}\, ,\quad b \equiv \frac{1-W}{\sqrt{1-W^2}} = \sqrt{\frac{1-W}{1+W}}\, ,
    \end{equation}
    and we have assumed that
    \begin{displaymath}
      \bar{z} = -z\, ,\quad \bar{\sigma} = -\sigma\, .
    \end{displaymath}
    Therefore, up to some constant the 1-loop determinant for the matter sector in the general background is
    \begin{equation}\label{eq:ZmatterSector}
      Z_{\textrm{mat}}^{1-\textrm{loop}} = \prod_\rho s_b \left(\frac{Q}{2} \left(\frac{z - q\rho(\sigma)}{\frac{v}{\ell_1}} - ir + i \right)\right)\, ,
    \end{equation}
    where
    \begin{equation}\label{eq:def-Q}
      Q \equiv b + b^{-1}\, ,\quad b \equiv \frac{1 - W}{\sqrt{1-W^2}} = \sqrt{\frac{1-W}{1+W}}\, ,\quad W \equiv \frac{\frac{v}{\ell} - \frac{v}{\ell_1}}{\frac{v}{\ell_1}}\, ,
    \end{equation}
    and $s_b(x)$ is the double-sine function, whose properties are discussed in Ref.~\cite{Koyama} and Appendix A of Ref.~\cite{Pasquetti-2}. For the general background auxiliary fields \eqref{eq:genAux} there is 
    \begin{equation}\label{eq:vl1}
      \frac{v}{\ell_1} = A_1 - \frac{1}{2} V_1 + iH = \frac{v}{\ell} \left(1 - \frac{2i}{v^2}\, \textrm{sin}\Theta\, e^{-i\Theta} \right)\, ,
    \end{equation}
    which leads to
    \begin{equation}
      W = \frac{2}{-2 + v^2 - i v^2 \textrm{cot}\Theta}\, ,\quad b = \sqrt{1 - \frac{2}{v^2} (1 - e^{-2i\Theta})}\, .
    \end{equation}
    As a quick check, we see that for the round $S^3$ there is $v = 1$ and $\Theta = 0$, then as expected
    \begin{displaymath}
      W = 0\, ,\quad b = 1\, .
    \end{displaymath}
    Moreover, for $\Theta=0$ there is always $b=1$, and by choosing different $v$ and $\ell$ in $\frac{v}{\ell_1}$ \eqref{eq:vl1} one obtains the result for round $S^3$ \cite{Kapustin-1} and the result in Ref.~\cite{Jap-2}. If we choose $\ell$ and $\Theta$ to be the ones given in Eqs.~\eqref{eq:IYsol-1}-\eqref{eq:IYsol-3}, and use the following identity for the double-sine function
    \begin{equation}
      s_b(x) \, s_b (-x) = 1\, ,
    \end{equation}
    we obtain the result in Ref.~\cite{IY}. Hence, the result \eqref{eq:ZmatterSector} incorporates all the previous results for a squashed $S^3$ with $SU(2)\times U(1)$ isometry.

    \subsubsection{1-Loop Determinant for Gauge Sector}
    In this section we discuss the 1-loop determinant for the gauge sector. The method is similar to the matter sector. We will see how the modes are paired, and how the missing spinors and the unpaired spinors give rise to the 1-loop determinant. First, the Lagrangians that are used to do the localization in the gauge sector, also have the bosonic part \eqref{eq:gaugeLag-1} and the fermionic part \eqref{eq:gaugeLag-2}. We can rescale the fields in the vector multiplet appropriately, then the Lagrangians \eqref{eq:gaugeLag-1} \eqref{eq:gaugeLag-2} become
    \begin{align}
      (Q \mathcal{V}_g)_B & = \textrm{Tr} \left[-\frac{1}{2} f_{\mu\nu} f^{\mu\nu} - (D_\mu \sigma) (D^\mu \sigma) + (D + \sigma H)^2 \right]\, , \label{eq:gaugeLagnew-1}\\
      (Q \mathcal{V}_g)_F & = \textrm{Tr} \left[2i \widetilde{\lambda} \slashed D \lambda + 2i [\widetilde{\lambda},\, \sigma] - iH (\widetilde{\lambda} \lambda) - V_1 (\widetilde{\lambda} \lambda) - 2V_\mu (\widetilde{\lambda} \gamma^\mu \lambda) \right]\, . \label{eq:gaugeLagnew-2}
    \end{align}

    We follow the same procedure as in Ref.~\cite{Kapustin-1}. First, we add a gauge fixing term:
    \begin{equation}
      \mathscr{L}_{gf} = \textrm{Tr} \left[\bar{c} \nabla^\mu \nabla_\mu c + b \nabla^\mu a_\mu \right]\, .
    \end{equation}
    Integration over $b$ will give the gauge fixing condition
    \begin{equation}
      \nabla^\mu a_\mu = 0\, .
    \end{equation}
    If we decompose
    \begin{equation}
      a_\mu = \nabla_\mu \varphi + B_\mu
    \end{equation}
    with
    \begin{equation}
      \nabla^\mu B_\mu = 0\, ,
    \end{equation}
    then the gauge fixing condition becomes
    \begin{equation}
      \nabla^\mu A_\mu = 0\quad \Rightarrow \quad \nabla^\mu \nabla_\mu \varphi = 0\, ,
    \end{equation}
    which is equivalent to a $\delta$-function in the Lagrangian:
    \begin{equation}
      \delta(\nabla^\mu A_\mu) = \delta(\nabla^2 \varphi) = \frac{1}{\sqrt{\textrm{det} (\nabla^2)}}\, \delta(\varphi)\, .
    \end{equation}
    After Gaussian integration, $c$ and $\bar{c}$ will contribute a factor to the partition function:
    \begin{displaymath}
      \textrm{det} (\nabla^2)\, ,
    \end{displaymath}
    while integration over $\sigma$ will contribute another factor
    \begin{displaymath}
      \frac{1}{\sqrt{\textrm{det} (\nabla^2)}}\, .
    \end{displaymath}
    Hence, the contributions from $c$, $\bar{c}$, $\sigma$ and $\varphi$ will cancel each other. Hence, around the classical solution \eqref{eq:ClassSol}, the bosonic part that contributes to the 1-loop determinant for the gauge sector, becomes
    \begin{equation}\label{eq:gaugeLagnew-3}
      (Q \mathcal{V}_g)'_B = \textrm{Tr} \left(B^\mu \hat{\Delta}_B B_\mu + [B_\mu,\, \sigma]^2 \right)\, ,
    \end{equation}
    where $\hat{\Delta}_B B_\mu \equiv *d *d B_\mu$.

    We see that the Lagrangians \eqref{eq:gaugeLagnew-2} and \eqref{eq:gaugeLagnew-3} are exactly the same as the ones in Ref.~\cite{MS-new}, therefore, we follow the same way as Ref.~\cite{MS-new} to figure out the pairing of modes. Suppose that there is a fermionic mode $\Lambda$ satisfying
    \begin{align}\label{eq:DeltalambdaLambda}
      {} & \Delta_\lambda \Lambda \equiv \Omega (i \gamma^\mu D_\mu + i\sigma \alpha - \frac{i}{2} H - \frac{1}{2} V_1 - V_\mu \gamma^\mu) \Lambda = M \Lambda\nonumber\\
      \Rightarrow \quad & (iM + \sigma \alpha \Omega) \Lambda = \Omega (-\gamma^\mu D_\mu + \frac{1}{2} H - \frac{i}{2} V_1 - iV_\mu \gamma^\mu) \Lambda\, .
    \end{align}
    Similar to Ref.~\cite{MS-new}, we can use the equation above to prove the following important relation by direct computations:
    \begin{equation}\label{eq:defB}
      B \equiv \Omega d (\widetilde{\zeta} \Lambda) + (iM + \sigma \alpha \Omega) (\widetilde{\zeta} \gamma_\mu \Lambda) d \xi^\mu = -i * \left(D(\widetilde{\zeta} \gamma_\mu \Lambda) d\xi^\mu \right)\, .
    \end{equation}
    In our case $\Omega$ is a constant, and the details of deriving this relation are given in Appendix E. Then this relation leads to
    \begin{align}
      {} & B \equiv \Omega d(\widetilde{\zeta} \Lambda) + (iM + \sigma \alpha \Omega) (\widetilde{\zeta} \gamma_\mu \Lambda) d\xi^\mu = -i * d \left(\frac{B - \Omega d(\widetilde{\zeta} \Lambda)}{iM + \sigma \alpha \Omega} \right) \nonumber\\
      \Rightarrow \quad & (iM + \sigma \alpha \Omega) B = -i * dB \label{eq:GaugeSectorTemp-1}\\
      \Rightarrow \quad & \pm \sqrt{\hat{\Delta}_B}\,\, B = *dB = -(M - i\sigma \alpha \Omega) B = -(M - i\sigma_0 \alpha) B\, ,\label{eq:importantRel}
    \end{align}
    with $\sigma_0 \equiv \sigma \Omega$. I.e., if there exists a fermionic mode $\Lambda$ with eigenvalue $M$ for $\Delta_\lambda$, then there is a corresponding bosonic mode with eigenvalue $-(M - i \alpha(\sigma_0))$ for $*d$. Conversely, if the relation \eqref{eq:importantRel} is true, i.e., there is a bosonic mode with eigenvalue $-(M - i \alpha(\sigma_0))$ for $*d$, then the fermionic mode $\Lambda \equiv \gamma^\mu B_\mu \zeta$ satisfies
    \begin{equation}
      \Delta_\lambda \Lambda = M \Lambda\, .
    \end{equation}
    In other words, the eigenmodes of $\sqrt{\hat{\Delta}_B}$ with eigenvalues $\pm (M - i \alpha(\sigma_0))$ are paired with the eigenmodes of $\Delta_\lambda$ with eigenvalues $M,\, -M + 2i \alpha(\sigma_0)$. On these paired bosonic modes $\Delta_B \equiv \hat{\Delta}_B + \alpha^2 (\sigma_0)$ has the eigenvalue $M (M - 2i \alpha (\sigma_0))$. Hence,
    \begin{equation}
      \frac{\Delta_\lambda}{\Delta_B} = \frac{-M (M - 2i\alpha(\sigma_0))}{M (M - 2i \alpha(\sigma_0))}\, .
    \end{equation}
    We see that up to some constant the paired modes cancel out exactly and do not have net contributions to the 1-loop determinant.

    Similar to the matter sector, to calculate the 1-loop determinant of the gauge sector, we still consider the unpaired spinor and the missing spinor. For the unpaired spinor, there is no corresponding bosonic mode, i.e.,
    \begin{equation}\label{eq:unpairedSpGaugeSector}
      \Omega \partial_\mu (\widetilde{\zeta} \Lambda) + (iM + \alpha(\sigma_0)) \widetilde{\zeta} \gamma_\mu \Lambda = 0\, ,
    \end{equation}
    and now the fermionic mode $\Lambda$ is
    \begin{equation}
      \Lambda = \zeta \Phi_0 + \zeta^c \Phi_2\, ,
    \end{equation}
    where $\Phi_0$ and $\Phi_2$ are bosonic fields with $R$-charges $0$ and $2$ respectively. Using the convention in Appendix A, we obtain for the first component of Eq.~\eqref{eq:unpairedSpGaugeSector}:
    \begin{align}\label{eq:MLambda}
      {} & \Omega^2 \partial_1 \Phi_0 + \Omega (iM_\Lambda + \alpha(\sigma_0)) \Phi_0 = 0 \nonumber\\
      \Rightarrow\quad & M_\Lambda = i \alpha(\sigma_0) + i \Omega \partial_1 = i \Omega \alpha(\sigma) + i \Omega \partial_1\, .
    \end{align}
    For the missing spinor there is
    \begin{equation}
      \Lambda = \gamma^\mu B_\mu \zeta = 0\, .
    \end{equation}
    By multiplying $\zeta^\dagger$ and $\zeta^{c\dagger}$ from the left, we see that this relation implies that
    \begin{displaymath}
      B_1 = 0\, ,\quad B_2 + iB_3 = 0\,\, \Rightarrow \,\, B_3 = iB_2\, .
    \end{displaymath}
    Then
    \begin{equation}\label{eq:GaugeSectorTemp-2}
      *dB = (i \partial_2 B_2 - \partial_3 B_2) e^1 - i\Omega \partial_1 B_2 e^2 + \Omega \partial_1 B_2 e^3\, .
    \end{equation}
    Combining Eq.~\eqref{eq:GaugeSectorTemp-1} and Eq.~\eqref{eq:GaugeSectorTemp-2}, we obtain
    \begin{align}
      {} & -i \Omega \partial_1 B_2 = - (M - i\alpha(\sigma_0)) iB_2 \nonumber\\
      \Rightarrow\quad & M_B = i\alpha(\sigma_0) + i \Omega \partial_1 = i \Omega \alpha(\sigma) + i \Omega \partial_1\, . \label{eq:MB}
    \end{align}
    Actually, as discussed in Refs.~\cite{Jap-2, MS-new}, there are more conditions that can be deduced, but for the case studied in this paper, the additional conditions are irrelevant. Moreover, it seems that the contributions from the unpaired spinors and the missing spinors \eqref{eq:MLambda} \eqref{eq:MB} cancel each other exactly. This is not the case, because the derivative $\partial_1$ acting on $\Phi_0$ and $B_2$ gives the eigenvalues proportional to the third component of the orbital angular momentum, i.e. $L_1$, while $\Phi_0$ and $B_2$ have spin $0$ and $1$ respectively. To know the precise form of the eigenvalues of $\partial_1$, i.e. the covariant derivative without spin and gauge connections, we use the same expression obtained in the matter sector \eqref{eq:nablaAfterTwist}:
    \begin{equation}
      \partial_1 = \frac{v}{\ell_1} 2 L^1 + (\frac{v}{\ell} - \frac{v}{\ell_1}) 2 L'^1 = \frac{v}{\ell_1} 2 (im - S_1) + (\frac{v}{\ell} - \frac{v}{\ell_1}) 2 im'\, ,
    \end{equation}
    where for the mode $\Phi_0$ the eigenvalue of $S_1$ is $0$, while for the mode $B_2$ the eigenvalue of $S_1$ can be $+1$ or $-1$, but since $\alpha$ runs over all the positive roots and negative roots, and $m$ and $m'$ run from $-j$ to $j$, the sign is actually irrelevant.

Considering all the modes contributing to $\frac{M_\Lambda}{M_B}$ in the most general background \eqref{eq:genAux}, we obtain the 1-loop determinant for the gauge sector:
    \begin{align}
      Z_g^{1-\textrm{loop}} & = \prod_{\alpha \in \Delta} \, \prod_j \prod_{-j \leqslant m,\, m' \leqslant j} \frac{-2 \frac{v}{\ell_1} m - 2m' \left(\frac{v}{\ell} - \frac{v}{\ell_1} \right) + i\alpha(\sigma)}{-2 \frac{v}{\ell_1} (m-1) - 2m' \left(\frac{v}{\ell} - \frac{v}{\ell_1} \right) + i\alpha(\sigma)} \nonumber\\
      {} & = \prod_{\alpha \in \Delta} \, \prod_j \prod_{-j \leqslant m' \leqslant j} \frac{-2 \frac{v}{\ell_1} j - 2m' \left(\frac{v}{\ell} - \frac{v}{\ell_1} \right) + i\alpha(\sigma)}{-2 \frac{v}{\ell_1} (-j-1) - 2m' \left(\frac{v}{\ell} - \frac{v}{\ell_1} \right) + i\alpha(\sigma)} \nonumber\\
      {} & = \prod_{\alpha \in \Delta} \, \prod_j \prod_{-j \leqslant m' \leqslant j} \left(- \frac{2j + 2m' W - \frac{i\alpha(\sigma)}{\frac{v}{\ell_1}}}{2j + 2 - 2m' W + \frac{i\alpha(\sigma)}{\frac{v}{\ell_1}}} \right) \nonumber\\
      {} & = \prod_{\alpha \in \Delta} \, \prod_{p,\, q = 0}^\infty \left(- \frac{p+q + (p-q) W - \frac{i\alpha(\sigma)}{\frac{v}{\ell_1}}}{p+q + 2 - (p-q) W + \frac{i\alpha(\sigma)}{\frac{v}{\ell_1}}} \right) \nonumber\\
      {} & = \prod_{\alpha \in \Delta} \, \prod_{p,\, q = 0}^\infty \left(- \frac{(1+W) p + (1-W) q + 1 - i \left(\frac{i\alpha(\sigma)}{\frac{v}{\ell_1}} - i \right)}{(1-W) p + (1+W) q + 1 + i\left(\frac{i\alpha(\sigma)}{\frac{v}{\ell_1}} - i \right) } \right) \nonumber\\
      {} & = \prod_{\alpha \in \Delta} \, \prod_{p,\, q = 0}^\infty \left(- \frac{b^{-1} p + b q + \frac{Q}{2} - \frac{iQ}{2} \left(\frac{i\alpha(\sigma)}{\frac{v}{\ell_1}} - i \right)}{b p + b^{-1} q + \frac{Q}{2} + \frac{iQ}{2} \left(\frac{i\alpha(\sigma)}{\frac{v}{\ell_1}} - i \right)} \right)\, ,
    \end{align}
    where $\alpha$ denotes the roots, and $p$, $q$, $W$, $b$ and $Q$ are defined in the same way as before \eqref{eq:def-jm} \eqref{eq:def-Wb} \eqref{eq:def-Q}. Up to some constant, the 1-loop determinant for the gauge sector can be written as
    \begin{equation}\label{eq:ZgaugeSector}
      Z_g^{1-\textrm{loop}} = \prod_{\alpha \in \Delta} s_b \left(\frac{Q}{2} \left(\frac{i\alpha(\sigma)}{\frac{v}{\ell_1}} - i \right) \right)\, ,
    \end{equation}
    where $s_b(x)$ is the double-sine function, and $\frac{v}{\ell_1}$ is given by Eq.~\eqref{eq:vl1}. By choosing appropriate parameters given by Eqs.~\eqref{eq:IYsol-1}-\eqref{eq:IYsol-3}, we obtain the result of Ref.~\cite{IY} from this general result.

    To see how the results of other cases emerge from the general one \eqref{eq:ZgaugeSector}, we need to rewrite the expression at some intermediate step. If we define
    \begin{equation}      b_1 + b_2 \equiv -2 \frac{v}{\ell_1}\, ,\quad b_1 - b_2 \equiv -2 \left(\frac{v}{\ell} - \frac{v}{\ell_1} \right)\, ,
    \end{equation}
    then
    \begin{align}
      Z_g^{1-\textrm{loop}} & = \prod_{\alpha \in \Delta} \, \prod_j \prod_{-j \leqslant m' \leqslant j} \frac{-2 \frac{v}{\ell_1} j - 2m' \left(\frac{v}{\ell} - \frac{v}{\ell_1} \right) + i\alpha(\sigma)}{-2 \frac{v}{\ell_1} (-j-1) - 2m' \left(\frac{v}{\ell} - \frac{v}{\ell_1} \right) + i\alpha(\sigma)} \nonumber\\
      {} & = \prod_{\alpha \in \Delta} \, \prod_j \prod_{-j \leqslant m' \leqslant j} \frac{-2 \frac{v}{\ell_1} j - 2m' \left(\frac{v}{\ell} - \frac{v}{\ell_1} \right) + i\alpha(\sigma)}{-2 \frac{v}{\ell_1} (-j-1) + 2m' \left(\frac{v}{\ell} - \frac{v}{\ell_1} \right) + i\alpha(\sigma)} \nonumber\\
      {} & = \prod_{\alpha \in \Delta} \, \prod_j \prod_{-j \leqslant m' \leqslant j} \frac{(b_1 + b_2) j + (b_1 - b_2) m' + i\alpha(\sigma) }{-(b_1 + b_2) (j+1) - (b_1 - b_2) m' + i\alpha(\sigma)} \nonumber\\
      {} & = \prod_{\alpha \in \Delta} \, \prod_j \prod_{-j \leqslant m' \leqslant j} \frac{b_1 (j+m') + b_2 (j-m') + i\alpha(\sigma)}{-b_1 (j+m'+1) - b_2 (j-m'+1) + i\alpha(\sigma)} \nonumber\\
      {} & = \prod_{\alpha \in \Delta} \, \prod_j \prod_{p,\, q = 0}^\infty \frac{b_1 p + b_2 q + i\alpha(\sigma)}{-b_1 (p+1) - b_2 (q+1) + i\alpha(\sigma)}\, ,
    \end{align}
    which is exactly the result of the 1-loop determinant for the gauge sector in Ref.~\cite{MS-new}. From this result, it is easy to obtain the results for other cases. To see it, we should rewrite the expression further and restrict $\alpha$ to be positive roots.
    \begin{align}
      Z_g^{1-\textrm{loop}} & = \prod_{\alpha\in \Delta_+}\, \prod_j \prod_{p,\, q = 0}^\infty \left(\frac{b_1 p + b_2 q + i\alpha(\sigma)}{-b_1 (p+1) - b_2 (q+1) + i\alpha(\sigma)} \cdot \frac{b_1 p + b_2 q - i\alpha(\sigma)}{-b_1 (p+1) - b_2 (q+1) - i\alpha(\sigma)} \right) \nonumber\\
      {} & = \prod_{\alpha\in \Delta_+}\, \prod_j \prod_{p,\, q = 0}^\infty \left(\frac{b_1 p + b_2 q + i\alpha(\sigma)}{-b_1 (p+1) - b_2 (q+1) + i\alpha(\sigma)} \cdot \frac{- b_1 p - b_2 q + i\alpha(\sigma)}{b_1 (p+1) + b_2 (q+1) + i\alpha(\sigma)} \right) \nonumber\\
      {} & = \prod_{\alpha\in \Delta_+}\, \left(\prod_{m>0} (b_1 m + i\alpha(\sigma))\cdot  \prod_{n>0} (b_2 n + i\alpha(\sigma))\cdot \prod_{m>0} (-b_1 m + i\alpha(\sigma))\cdot \prod_{n>0} (-b_2 n + i\alpha(\sigma))\right) \nonumber\\
      {} & = \prod_{\alpha\in \Delta_+}\, \prod_{m>0} (b_1^2 m^2 + \alpha^2 (\sigma)) \cdot \prod_{n>0} (b_2^2 n^2 + \alpha^2 (\sigma)) \nonumber\\
      {} & = \prod_{\alpha\in \Delta_+}\, \prod_{n>0} \left((b_1^2 n^2 + \alpha^2 (\sigma))\cdot (b_2^2 n^2 + \alpha^2 (\sigma)) \right)\, ,
    \end{align}
    where $\Delta_+$ denotes the set of positive roots. For
    \begin{displaymath}
      b_1 = b_2 = \frac{1}{\ell}\, ,
    \end{displaymath}
    we obtain the result of round $S^3$ \cite{Kapustin-1}, while for
    \begin{displaymath}
      b_1 = b_2 = \frac{v}{\ell} = \frac{1}{\tilde{\ell}}\, ,
    \end{displaymath}
    the result becomes the one of the squashed $S^3$ with $SU(2)\times U(1)$ isometry discussed in Ref.~\cite{Jap-2}. Hence, like in the matter sector, the general result \eqref{eq:ZgaugeSector} also incorporates all the previous results on a squashed $S^3$ with $SU(2)\times U(1)$ isometry, and it does not depend on the shifts by $\sim \kappa$ of the auxiliary fields, instead the shifts induced by the rotation of the Killing spinors will affect the final result.

    Finally, putting everything together \eqref{eq:Zclass-1}-\eqref{eq:Zclass-3} \eqref{eq:ZmatterSector} \eqref{eq:ZgaugeSector}, we obtain the results summarized in the introduction. As we emphasized there, an important feature is that the 1-loop determinants are independent of the shift $\kappa$, while only $\Theta \neq 0$ can give the results essentially different from the case of the round $S^3$.

\section{Conclusion and Discussion}
Through the calculations above, we have seen that Refs.~\cite{Seiberg, 4D-I, 4D-II, 3D} provide a very powerful tool for studying theories with rigid supersymmetry on curved space. It is especially useful in finding the Killing spinors and the corresponding background auxiliary fields, which could be extremely hard but at the same time crucial to the calculations in the localization procedure. These new formalisms circumvent this difficulty in a systematic way.

The partition function of the $\mathcal{N}=2$ supersymmetric Chern-Simons-Matter theory on a squashed $S^3$ with $SU(2)\times U(1)$ isometry and a class of complex background is calculated. The result provides a new interpolation between the results of Ref.~\cite{Jap-2} and Ref.~\cite{IY}. From this example we see clearly how the background fields enter the final results. Even for the round $S^3$, i.e. $v=1$, the partition function can be nontrivial if $\Theta \neq 0$. Moreover, different background fields do not necessarily lead to different partition functions. For instance, the shift proportional to $\kappa$ in the background auxiliary fields leaves the 1-loop determinants in the partition function unchanged. This work complements Ref.~\cite{MS-new} by studying a special case in great detail.

It would be interesting to see whether the same features are shared by other kinds of 3D and 4D manifolds, i.e., whether the partition functions for the theory defined on the manifolds depend on the background auxiliary fields in a certain way. Moreover, Ref.~\cite{gravdual} has studied the gravity dual for the Chern-Simons-Matter theory defined on a squashed $S^3$ with $SU(2)\times U(1)$ isometry. With the general results discussed in this paper, we would like to see how the large class of gauge theories fits into the gravity side, and how the auxiliary fields appear appropriately in the gravity theory.

\section*{Acknowledgments}
I would like to thank Francesco Benini, Marcos Crichigno, Chris Herzog, Nikita Nekrasov, Yiwen Pan, Wolfger Peelaers, Martin Ro\v cek, Xinyu Zhang and Peng Zhao for many useful discussions. I am also very grateful to Guido Festuccia, Yosuke Imamura and Dario Martelli for very patient and detailed answers to my questions. Special thanks to Kazuo Hosomichi and Sungjay Lee for enlightening discussions that help me resolve some key problems in the calculations. This work was supported by NSF grant No. PHY-0969739.

\appendix
\section{Convention}
  In this appendix we review our convention and some identities used in the paper. We mainly follow the convention of Ref.~\cite{3D}. The 3D $\gamma$-matrices are chosen to be
  \begin{equation}
    \gamma_1 = \sigma_3\, ,\quad \gamma_2 = -\sigma_1\, ,\quad \gamma_3 = -\sigma_2\, ,
  \end{equation}
  where $\sigma_i$ are the Pauli matrices. They still satisfy
  \begin{equation}
    [\gamma_m,\, \gamma_n] = 2i \varepsilon_{mnp} \gamma^p\, .
  \end{equation}
  This will consequently affect the eigenvalues of spherical harmonics defined on the squashed $S^3$. The main difference is that for a spin-0 field instead of $L_3$ now $L_1$ has the eigenvalues
  \begin{displaymath}
    im\quad \textrm{with} \quad -j \leqslant m \leqslant j,\quad j = 0,\, \frac{1}{2},\, 1,\, \cdots
  \end{displaymath}

  In this paper, we use commuting spinors. The product of two spinors are defined as
  \begin{equation}
    \psi \chi = \psi^\alpha C_{\alpha\beta} \chi^\beta\, ,\quad \psi \gamma_\mu \chi = \psi^\alpha (C \gamma_\mu)_{\alpha\beta} \chi^\beta\, ,
  \end{equation}
  where the indices can be raised and lowered using
  \begin{displaymath}
    C = \left(\begin{array}{cc}
                                        0 & 1\\
                                        -1 & 0
                                      \end{array} \right)
  \end{displaymath}
  is the charge conjugation matrix. The spinor bilinears of commuting spinors satisfy
  \begin{equation}
    \psi \chi = -\chi \psi\, ,\quad \psi \gamma_\mu \chi = \chi \gamma_\mu \psi\, .
  \end{equation}
  The Fierz identity for commuting spinors is
  \begin{equation}
    (\psi_1 \chi_1) (\psi_2 \chi_2) = \frac{1}{2} (\psi_1 \chi_2) (\psi_2 \chi_1) + \frac{1}{2} (\psi_1 \gamma^\mu \chi_2) (\psi_2 \gamma_\mu \chi_1)\, .
  \end{equation}
  The Hermitian conjugate of a spinor is given by
  \begin{equation}
    (\psi^\dagger)^\alpha \equiv \overline{(\psi_\alpha)}\, ,
  \end{equation}
  where $\overline{\phantom{a}}$ denotes the complex conjugate. The charge conjugate of a spinor is defined as
  \begin{equation}
    \chi^c \equiv \sigma_2 \overline{\chi}\, .
  \end{equation}

  We use $\psi$ and $\widetilde{\psi}$ to denote the spinors that are Hermitian conjugate to each other in Lorentzian signature, while independent in Euclidean signature. In particular, $\zeta$ and $\widetilde{\zeta}$ are such a kind of spinor pair. They satisfy the generalized Killing spinor equations \eqref{eq:genKilling}. Although in Euclidean signature they are independent, one can prove that the charge conjugate of $\zeta$, i.e., $\zeta^c$ satisfies the same Killing spinor equation as $\widetilde{\zeta}$. Hence,
  \begin{equation}
    \zeta^c \propto \widetilde{\zeta}\, .
  \end{equation}

\section{$S^3$ as an $SU(2)$-Group Manifold}
  A convenient way to discuss different ways of squashing is to introduce the left-invariant and the right-invariant frame. A group element in $SU(2)$ is given by
  \begin{equation}
    g \equiv i x_\mu \sigma^\mu = \left( \begin{array}{cc}
                   x_0 + i x_3 & x_2 + i x_1\\
                   -x_2 + i x_1 & x_0 - i x_3
                 \end{array}\right)\, ,
  \end{equation}
  where $\sigma^\mu = (-i I, \vec{\sigma})$. Then
  \begin{equation}
    \mu \equiv g^{-1} dg\quad \textrm{and}\quad \widetilde{\mu} \equiv dg\, g^{-1}
  \end{equation}
  are left-invariant and right-invariant 1-form respectively, i.e., they are invariant under the transformations with a constant matrix $h$
  \begin{displaymath}
    g \to h\, g\quad \textrm{and}\quad g \to g\, h
  \end{displaymath}
  respectively. It can be checked explicitly that the metric of $S^3$ can be written as
  \begin{equation}\label{eq:metricInForms}
    ds^2 = \frac{\ell^2}{2} \textrm{tr}(dg\, dg^{-1}) = \ell^2 \mu^m \mu^m = \ell^2 \widetilde{\mu}\,^m\, \widetilde{\mu}\,^m = \frac{\ell^2}{2} (\mu^m \mu^m + \widetilde{\mu}\,^m\, \widetilde{\mu}\,^m)\, ,
  \end{equation}
  with $m = 1,\, 2,\, 3$, if we impose the constraint
  \begin{equation}
    \mathrm{det}\, g = x_0\,^2 + x_1\,^2 + x_2\,^2 + x_3\,^2 = 1\, ,
  \end{equation}
  where
  \begin{equation}
    \mu^m = \frac{i}{2} \textrm{tr} (\mu \gamma^m)\, ,\quad \widetilde{\mu}\,^m = \frac{i}{2} \textrm{tr} (\widetilde{\mu}\, \gamma^m)\, ,
  \end{equation}
  or equivalently,
  \begin{equation}
    -2 \mu^m T_m = \mu = g^{-1} dg\, ,\quad T_m \equiv i \frac{\gamma_m}{2}\, ,\,\, g\in SU(2)\, .
  \end{equation}
  Hence, the metric of $S^3$ is both left-invariant and right-invariant.

  A 3D Killing spinor is a spinor that satisfies
  \begin{equation}
    D \epsilon \equiv d \epsilon + \frac{1}{4} \gamma_{mn} \omega^{mn} \epsilon = e^m \gamma_m \widetilde{\epsilon}\, ,
  \end{equation}
  where $\gamma_{mn} \equiv \frac{1}{2} (\gamma_m \gamma_n - \gamma_n \gamma_m)$, and the choice of $\gamma_m$ is given in Appendix A, while $\omega^{mn}$ is the spin connection, and $\widetilde{\epsilon}$ is another spinor which in general can be different from $\epsilon$. In this paper, sometimes we try to bring the spinors satisfying generalized Killing spinor equations back into this simple form. Moreover, we can define two Killing vector fields by their group actions on a general group element $g$ in $SU(2)$:
  \begin{equation}
    \mathcal{L}^m g = i\gamma^m g\, ,\qquad \mathcal{R}^m g = ig \gamma^m\, .
  \end{equation}
  As discussed before, the metric of $S^3$ is both left-invariant and right-invariant, hence it is also invariant under the actions of $\mathcal{L}^m$ and $\mathcal{R}^m$ given above. By definition, a Killing vector field is a vector field that preserves the metric. Therefore, $\mathcal{L}^m$ and $\mathcal{R}^m$ are Killing vector fields. Since the actions of $\mathcal{L}^m$ and $\mathcal{R}^m$ are equivalent to multiplications of $i\gamma^m$ from the left and from the right respectively, after some rescaling they are the same as the generators of $SU(2)$ algebra, i.e., $\{\frac{1}{2i} \mathcal{L}^m\}$ and $\{-\frac{1}{2i} \mathcal{R}^m\}$ both satisfy the commutation relation of the $SU(2)$ algebra.

  In the above, we define $\mathcal{L}^m$ and $\mathcal{R}^m$ as group actions. Sometimes we also use them to denote the variations caused by the group actions. In this sense, there are
  \begin{equation}
    \mathcal{L}^m (g^{-1} dg) = 0 \quad \Rightarrow \quad \mathcal{L}^m \mu^n = 0
  \end{equation}
  and
  \begin{align}
    \mathcal{L}^m \widetilde{\mu}\,^n & = \mathcal{L}^m \frac{i}{2} \textrm{tr}(\widetilde{\mu}\gamma^n) = \mathcal{L}^m \frac{i}{2} \textrm{tr}(dg \, g^{-1} \gamma^n)\nonumber\\
    {} & = \frac{i}{2} \left[\textrm{tr} (i\gamma^m \widetilde{\mu} \gamma^n) + \textrm{tr} (\widetilde{\mu} (-i \gamma^m) \gamma^n) \right] \nonumber\\
    {} & = 2\varepsilon^{mnp}\, \widetilde{\mu}\,^p\, .
  \end{align}
  Similarly, there are
  \begin{equation}
    \mathcal{R}^m\, \widetilde{\mu}\,^n = 0\quad \textrm{and}\quad \mathcal{R}^m \mu^n = -2\varepsilon^{mnp} \mu^p\, .
  \end{equation}
  In other words, $\mathcal{L}^m$ acts only on the right-invariant frames, while $\mathcal{R}^m$ acts only on the left-invariant frames. Hence, the round $S^3$ has an $SU(2)_L\times SU(2)_R$ isometry.

\section{Different Metrics of Squashed $S^3$}
  There are different expressions of $S^3$ and squashed $S^3$. In this appendix we review some relevant ones for this paper. A more thorough discussion can be found in Appendix A of Ref.~\cite{Okuda}.

  As discussed in Appendix B, the metric of $S^3$ can be written as \eqref{eq:metricInForms}:
  \begin{equation}
    ds^2 = \ell^2 \mu^m \mu^m = \ell^2 \widetilde{\mu}\,^m \widetilde{\mu}\,^m\, ,
  \end{equation}
  where
  \begin{equation}
    g^{-1} dg = -i \mu^m \gamma^m\, ,\quad dg'\, g'^{-1} = -i \widetilde{\mu}\,^m \gamma^m\, .
  \end{equation}
  In general the $SU(2)$ group elements $g$ and $g'$ can be different. If we choose on $S^3$
  \begin{displaymath}
    g = \left( \begin{array}{cc}
                 \textrm{cos} \left(\frac{\theta}{2} \right)\cdot \textrm{exp} \left(i \frac{\phi + \psi}{2} \right) & \textrm{sin} \left(\frac{\theta}{2} \right)\cdot \textrm{exp} \left(i \frac{\phi - \psi}{2} \right)\\
                 -\textrm{sin} \left(\frac{\theta}{2} \right)\cdot \textrm{exp} \left(-i \frac{\phi - \psi}{2} \right) & \textrm{cos} \left(\frac{\theta}{2} \right)\cdot \textrm{exp} \left(-i \frac{\phi + \psi}{2} \right)
               \end{array}\right)\, ,
  \end{displaymath}
  \begin{displaymath}
    g' = \left( \begin{array}{cc}
                  \textrm{cos} \left(\frac{\theta}{2} \right)\cdot \textrm{exp} \left(i \frac{\phi + \psi}{2} \right) & \textrm{sin} \left(\frac{\theta}{2} \right)\cdot \textrm{exp} \left(-i \frac{\phi - \psi}{2} \right)\\
                  -\textrm{sin} \left(\frac{\theta}{2} \right)\cdot \textrm{exp} \left(i \frac{\phi - \psi}{2} \right) & \textrm{cos} \left(\frac{\theta}{2} \right)\cdot \textrm{exp} \left(-i \frac{\phi + \psi}{2} \right)
                \end{array}\right)\, ,
  \end{displaymath}
  then the vielbeins in the left-invariant frame and in the right-invariant frame are given by:
  \begin{align}\label{eq:leftinvFrame}
    e_1^{(0)} & \equiv \ell \mu^1 = -\frac{\ell}{2} (d\psi + \textrm{cos} \theta\, d\phi)\, , \nonumber\\
    e_2^{(0)} & \equiv \ell \mu^2 = -\frac{\ell}{2} (\textrm{sin} \psi\, d\theta - \textrm{sin} \theta\, \textrm{cos} \psi\, d\phi)\, , \nonumber\\
    e_3^{(0)} & \equiv \ell \mu^3 = \frac{\ell}{2} (\textrm{cos} \psi\, d\theta + \textrm{sin} \theta\, \textrm{sin} \psi\, d\phi )\, .
  \end{align}
  \begin{align}\label{eq:rightinvFrame}
    \widetilde{e}_1^{(0)} & \equiv \ell\, \widetilde{\mu}\,^1 = -\frac{\ell}{2} (d\psi + \textrm{cos} \theta\, d\phi)\, , \nonumber\\
    \widetilde{e}_2^{(0)} & \equiv \ell\, \widetilde{\mu}\,^2 = \frac{\ell}{2} (\textrm{sin} \psi\, d\theta - \textrm{sin} \theta\, \textrm{cos} \psi\, d\phi)\, , \nonumber\\
    \widetilde{e}_3^{(0)} & \equiv \ell\, \widetilde{\mu}\,^3 = \frac{\ell}{2} (\textrm{cos} \psi\, d\theta + \textrm{sin} \theta\, \textrm{sin} \psi\, d\phi )\, .
  \end{align}
  They satisfy
  \begin{equation}
    d e^a_{(0)} + \frac{1}{\ell} \varepsilon^{abc} e^b_{(0)} \wedge e^c_{(0)} = 0 \, ,\quad d\, \widetilde{e}\,^a_{(0)} - \frac{1}{\ell} \varepsilon^{abc}\, \widetilde{e}\,^b_{(0)} \wedge \widetilde{e}\,^c_{(0)} = 0\, ,
  \end{equation}
  i.e., the spin connections are
  \begin{equation}
    \omega^{ab}_{(0)} = -\frac{1}{\ell} \varepsilon^{abc} e^c_{(0)} \, ,\quad \widetilde{\omega}\,^{ab}_{(0)} = \frac{1}{\ell} \varepsilon^{abc} \widetilde{e}\,^c_{(0)}\, .
  \end{equation}
  We see explicitly that for this choice of $g$ and $g'$ there are
  \begin{equation}\label{eq:e1Eqet1}
    e^1_{(0)} = \widetilde{e}\,^1_{(0)} = -\frac{\ell}{2} (d\psi + \textrm{cos} \theta\, d\phi)\, ,\quad (e^2_{(0)})^2 + (e^3_{(0)})^2 = (\widetilde{e}\,^2_{(0)})^2 + (\widetilde{e}\,^3_{(0)})^2 = \frac{\ell^2}{4} (d\theta^2 + \textrm{sin}^2 \theta d\phi^2)\, .
  \end{equation}
  That is the reason why in the paper we can occasionally interchange between the left-invariant frame and the right-invariant frame.

  Besides the form \eqref{eq:metricInForms}, the metric of $S^3$ can also be written as a Hopf fibration or a torus fibration. Both the left-invariant frame \eqref{eq:leftinvFrame} and the right-invariant frame \eqref{eq:rightinvFrame} can give the Hopf fibration of $S^3$:
  \begin{equation}\label{eq:metricHopf}
    ds^2 = \frac{\ell^2}{4} (d\theta^2 + \textrm{sin}^2 \theta d\phi^2 + (d\psi + \textrm{cos} \theta \, d\phi)^2)\, ,
  \end{equation}
  where
  \begin{equation}
    0 \leqslant \theta \leqslant \pi\, ,\quad 0 \leqslant \phi \leqslant 2\pi\, ,\quad 0 \leqslant \psi \leqslant 4\pi\, .
  \end{equation}
  The torus fibration of $S^3$ is
  \begin{equation}\label{eq:metricTorus}
    ds^2 = \ell^2 (d\vartheta^2 + \textrm{sin}^2 \vartheta d\varphi_1^2 + \textrm{cos}^2 \vartheta d\varphi_2^2)\, ,
  \end{equation}
  where
  \begin{equation}
    0 \leqslant \vartheta \leqslant \frac{\pi}{2}\, ,\quad 0 \leqslant \varphi_1\, ,\varphi_2 \leqslant 2\pi\, .
  \end{equation}
  The following conditions relate the coordiantes of the Hopf fibration and the torus fibration of $S^3$:
  \begin{equation}
    \theta = 2\vartheta\, ,\quad \phi = \varphi_2 - \varphi_1\, ,\quad \psi = \varphi_1 + \varphi_2\, .
  \end{equation}

  To apply the method introduced in Ref.~\cite{3D}, we have to rewrite the metric of $S^3$ as a Hopf fibration \eqref{eq:metricHopf} further. Using the stereographic projection
  \begin{equation}\label{eq:stereoCoord}
    X \equiv \textrm{cot} \frac{\theta}{2}\, \textrm{cos}\phi\, ,\quad Y \equiv \textrm{cot} \frac{\theta}{2}\, \textrm{sin}\phi
  \end{equation}
  we can rewrite the metric of $S^2$ as
  \begin{align}
    ds^2 & = d\theta^2 + \textrm{sin}^2 \theta d\phi^2 \nonumber\\
    {} & = \frac{4}{(1 + X^2 + Y^2)^2} (dX^2 + dY^2) \nonumber\\
    {} & = \frac{4}{(1 + z \bar{z})^2} dz \, d\bar{z}\, ,
  \end{align}
  where
  \begin{equation}\label{eq:complexCoord}
    z \equiv X + i Y\, ,\quad \bar{z} \equiv X - i Y\, .
  \end{equation}
  Consequently, the metric of $S^3$ as a Hopf fibration \eqref{eq:metricHopf} has the following forms, if we set $\ell = 1$:
  \begin{align}
    ds^2 & = \frac{1}{4} (d\psi + \textrm{cos}\theta\, d\phi)^2 + \frac{1}{4} d\theta^2 + \frac{1}{4} \textrm{sin}^2 \theta\, d\phi^2 \nonumber\\
    {} & = \frac{1}{4} \left[d\psi - \frac{X^2+Y^2-1}{X^2+Y^2+1} \cdot \frac{Y}{X^2+Y^2}\, dX + \frac{X^2+Y^2-1}{X^2+Y^2+1} \cdot \frac{X}{X^2+Y^2}\, dY \right]^2 \nonumber\\
    {} & \quad + \frac{1}{(1+X^2+Y^2)^2} (dX^2 + dY^2) \nonumber\\
    {} & = \frac{1}{4} (d\psi + a dz + \bar{a} d \bar{z})^2 + c^2 dz\, d\bar{z}\, ,
  \end{align}
  where
  \begin{equation}
    a \equiv -\frac{i}{2z} \cdot \frac{z\bar{z}-1}{z\bar{z}+1}\, ,\quad c \equiv \frac{1}{1+z\bar{z}}\, .
  \end{equation}

  Based on our choice of left-invariant frame \eqref{eq:leftinvFrame} and the right-invariant frame \eqref{eq:rightinvFrame}, the metric of the squashed $S^3$ with $SU(2)\times U(1)$ isometry has the following expression:
  \begin{equation}\label{eq:squashedMetricInForms}
    ds^2 = \frac{\ell^2}{v^2} \mu^1 \mu^1 + \ell^2 \mu^2 \mu^2 + \ell^2 \mu^3 \mu^3 = \frac{\ell^2}{v^2} \widetilde{\mu}\,^1 \widetilde{\mu}\,^1 + \ell^2 \widetilde{\mu}\,^2 \widetilde{\mu}\,^2 + \ell^2 \widetilde{\mu}\,^3 \widetilde{\mu}\,^3\, ,
  \end{equation}
  where $v$ is a constant squashing parameter. For this squashed $S^3$, we choose the left-invariant frame and the right-invariant frame to be
  \begin{align}\label{eq:sqLeftinvFrame}
    e_1 & \equiv \frac{\ell}{v} \mu^1 = -\frac{\ell}{2v} (d\psi + \textrm{cos} \theta\, d\phi)\, , \nonumber\\
    e_2 & \equiv \ell \mu^2 = -\frac{\ell}{2} (\textrm{sin} \psi\, d\theta - \textrm{sin} \theta\, \textrm{cos} \psi\, d\phi)\, , \nonumber\\
    e_3 & \equiv \ell \mu^3 = \frac{\ell}{2} (\textrm{cos} \psi\, d\theta + \textrm{sin} \theta\, \textrm{sin} \psi\, d\phi )\, .
  \end{align}
  \begin{align}\label{eq:sqRightinvFrame}
    \widetilde{e}_1 & \equiv \frac{\ell}{v}\, \widetilde{\mu}\,^1 = -\frac{\ell}{2v} (d\psi + \textrm{cos} \theta\, d\phi)\, , \nonumber\\
    \widetilde{e}_2 & \equiv \ell\, \widetilde{\mu}\,^2 = \frac{\ell}{2} (\textrm{sin} \psi\, d\theta - \textrm{sin} \theta\, \textrm{cos} \psi\, d\phi)\, , \nonumber\\
    \widetilde{e}_3 & \equiv \ell\, \widetilde{\mu}\,^3 = \frac{\ell}{2} (\textrm{cos} \psi\, d\theta + \textrm{sin} \theta\, \textrm{sin} \psi\, d\phi )\, .
  \end{align}
  The corresponding spin connections are
  \begin{equation}\label{eq:leftSpConn}
    \omega^{23} = -(2 - \frac{1}{v^2})\, \mu^1\, ,\quad \omega^{31} = -\frac{1}{v}\, \mu^2\, ,\quad \omega^{12} = -\frac{1}{v}\, \mu^3
  \end{equation}
  for the left-invariant frame, and
  \begin{equation}\label{eq:rightSpConn}
    \widetilde{\omega}\,^{23} = (2 - \frac{1}{v^2})\, \widetilde{\mu}\,^1\, ,\quad \widetilde{\omega}\,^{31} = \frac{1}{v}\, \widetilde{\mu}\,^2\, ,\quad \widetilde{\omega}\,^{12} = \frac{1}{v}\, \widetilde{\mu}\,^3
  \end{equation}
  for the right-invariant frame. As for $S^3$, the metric \eqref{eq:squashedMetricInForms} can also be written in some other coordinates:
  \begin{align}
    ds^2 & = \frac{1}{4v^2} (d\psi + \textrm{cos}\theta\, d\phi)^2 + \frac{1}{4} d\theta^2 + \frac{1}{4} \textrm{sin}^2 \theta\, d\phi^2 \label{eq:metric-thetaphipsi}\\
    {} & = \frac{1}{4v^2} \left[d\psi - \frac{X^2+Y^2-1}{X^2+Y^2+1} \cdot \frac{Y}{X^2+Y^2}\, dX + \frac{X^2+Y^2-1}{X^2+Y^2+1} \cdot \frac{X}{X^2+Y^2}\, dY \right]^2 \nonumber\\
    {} & \quad + \frac{1}{(1+X^2+Y^2)^2} (dX^2 + dY^2) \label{eq:metric-XYpsi}\\
    {} & = \frac{1}{4v^2} (d\psi + a dz + \bar{a} d \bar{z})^2 + c^2 dz\, d\bar{z}\, ,\label{eq:metric-complex}
  \end{align}
  In practice it is more convenient to choose a frame different from the right-invariant frame \eqref{eq:sqRightinvFrame}, which is given by
  \begin{align}
    \hat{e}^1 & = \frac{1}{2v} \left[d\psi - \frac{X^2+Y^2-1}{X^2+Y^2+1} \cdot \frac{Y}{X^2+Y^2}\, dX + \frac{X^2+Y^2-1}{X^2+Y^2+1} \cdot \frac{X}{X^2+Y^2}\, dY \right]\, , \nonumber\\
    \hat{e}^2 & = \frac{1}{1+X^2+Y^2} dX\, , \nonumber\\
    \hat{e}^3 & = \frac{1}{1+X^2+Y^2} dY\, .\label{eq:workingFrame}
  \end{align}
  In the coordinates $(\theta,\, \phi,\, \psi)$ and $(z,\, \bar{z},\, \psi)$ the vielbeins look like
  \begin{align}
    \hat{e}^1 & = \frac{1}{2v} d\psi + \frac{1}{2v} \textrm{cos} \theta\, d\phi = \frac{1}{2v} (d\psi + a dz + \bar{a} d\bar{z})\, ,\nonumber\\
    \hat{e}^2 & = -\frac{1}{2} \textrm{cos} \phi\, d\theta - \frac{1}{2} \textrm{sin} \theta\, \textrm{sin} \phi\, d\phi = c \frac{dz + d\bar{z}}{2}\, ,\nonumber\\
    \hat{e}^3 & = -\frac{1}{2} \textrm{sin} \phi\, d\theta + \frac{1}{2} \textrm{sin} \theta\, \textrm{cos} \phi\, d\phi = c \frac{dz - d\bar{z}}{2i}\, .\label{eq:workingFrame-2}
  \end{align}

\section{BPS Equations and Classical Solutions}
As we discussed in the text, to preserve the supersymmetry given by Eq.~\eqref{eq:fullSUSYgauge} and Eq.~\eqref{eq:fullSUSYmatter}, the BPS equations \eqref{eq:LocalCond} should be satisfied:
    \begin{equation}
      Q \psi = 0\, ,\quad Q \widetilde{\psi} = 0\, ,\quad Q \lambda = 0\, ,\quad Q \widetilde{\lambda} = 0\, ,
    \end{equation}
or in explicit form
\begin{align}
  0 & = \sqrt{2} \zeta F - \sqrt{2} i (z - q\sigma - r H) \widetilde{\zeta} \phi - \sqrt{2} i \gamma^\mu \widetilde{\zeta} D_\mu \phi\, ,\label{eq:BPS-1}\\
  0 & = \sqrt{2} \widetilde{\zeta} \widetilde{F} + \sqrt{2} i (z - q\sigma - r H) \zeta \widetilde{\phi} + \sqrt{2} i \gamma^\mu \zeta D_\mu \widetilde{\phi}\, ,\label{eq:BPS-2}\\
  0 & = i \zeta (D + \sigma H) - \frac{i}{2} \varepsilon^{\mu\nu\rho} \gamma_\rho \zeta f_{\mu\nu} - \gamma^\mu \zeta (i \partial_\mu \sigma - V_\mu \sigma)\, ,\label{eq:BPS-3}\\
  0 & = -i \widetilde{\zeta} (D + \sigma H) - \frac{i}{2} \varepsilon^{\mu\nu\rho} \gamma_\rho \widetilde{\zeta} f_{\mu\nu} + \gamma^\mu \widetilde{\zeta} (i \partial_\mu \sigma + V_\mu \sigma)\, .\label{eq:BPS-4}
\end{align}
Using the solutions of the Killing spinor equations \eqref{eq:KillingSpUp}
  \begin{equation}
    \zeta^\alpha = \sqrt{s} \left(\begin{array}{c} 0 \\ -1 \end{array} \right) \, ,\quad \widetilde{\zeta}^\alpha = \frac{1}{2v\sqrt{s}} \left(\begin{array}{c} 1 \\ 0 \end{array} \right)\, .
  \end{equation}
and choosing the frame given by \eqref{eq:workingFrame} \eqref{eq:workingFrame-2}, we obtain for commuting spinors
\begin{equation}
  \zeta \zeta = 0\, ,\quad \widetilde{\zeta} \widetilde{\zeta} = 0\, ,\quad \zeta \widetilde{\zeta} = - \widetilde{\zeta} \zeta = -\frac{1}{2v}\, ,
\end{equation}
\begin{equation}
  \zeta \gamma_m \zeta = (0, s , -i s)\, ,\quad \widetilde{\zeta} \gamma_m \widetilde{\zeta} = (0, -\frac{1}{4 s v^2}, -\frac{i}{4 s v^2})\, ,
\end{equation}
\begin{equation}
  \zeta \gamma_m \widetilde{\zeta} = (\frac{1}{2v}, 0 , 0)\, ,\quad \widetilde{\zeta} \gamma_m \zeta = (\frac{1}{2v}, 0 , 0)\, ,
\end{equation}
where $m = 1, 2, 3$.

Contracting Eq.~\eqref{eq:BPS-1} with $\widetilde{\zeta}$ and Eq.~\eqref{eq:BPS-2} with $\zeta$ from the left, one obtains
\begin{equation}
  F = 0\, ,\quad \widetilde{F} = 0\, .
\end{equation}
Plugging these solutions into Eq.~\eqref{eq:BPS-1} and Eq.~\eqref{eq:BPS-2}, can contracting them with $\zeta$ and $\widetilde{\zeta}$ respectively from the left, one has
\begin{align}
  \frac{\sqrt{2} i}{2v} (z - q\sigma - r H) \phi - \frac{\sqrt{2} i}{2v} D_1 \phi & = 0\, ,\\
  \frac{\sqrt{2} i}{2v} (z - q\sigma - r H) \widetilde{\phi} + \frac{\sqrt{2} i}{2v} D_1 \widetilde{\phi} & = 0\, .
\end{align}
Since $\widetilde{\phi} = \phi^\dagger$, for generic values of $(z - q\sigma - r H)$ the equations above do not have nontrivial solutions. Hence,
\begin{equation}
  \phi = \widetilde{\phi} = 0\, .
\end{equation}

  In the gauge sector, we can contract Eq.~\eqref{eq:BPS-3} and Eq.~\eqref{eq:BPS-4} with $\zeta$ and $\widetilde{\zeta}$ respectively from the left, then we obtain
  \begin{align}
    (\zeta \gamma_\mu \zeta) \left(-\frac{i}{2} \varepsilon^{\rho\sigma\mu} f_{\rho\sigma} - i \partial^\mu \sigma + V^\mu \sigma \right) & = 0\, ,\label{eq:GaugeBPS-1}\\
    (\widetilde{\zeta} \gamma_\mu \widetilde{\zeta}) \left(-\frac{i}{2} \varepsilon^{\rho\sigma\mu} f_{\rho\sigma} + i\partial^\mu \sigma + V^\mu \sigma \right) & = 0\, .\label{eq:GaugeBPS-2}
  \end{align}
  Taking into account that
  \begin{displaymath}
    \zeta \gamma_2 \zeta \neq 0\, ,\quad \zeta \gamma_3 \zeta \neq 0\, ,\quad \widetilde{\zeta} \gamma_2 \widetilde{\zeta} \neq 0\, ,\quad \widetilde{\zeta} \gamma_3 \widetilde{\zeta} \neq 0\, ,
  \end{displaymath}
  we obtain that
  \begin{equation}
    \partial_2 \sigma = \partial_3 \sigma = 0\, .
  \end{equation}
  Similarly, contracting  Eq.~\eqref{eq:BPS-3} and Eq.~\eqref{eq:BPS-4} with $\widetilde{\zeta}$ and $\zeta$ respectively from the left will give
  \begin{align}
    \frac{i}{2v} (D + \sigma H) + (\widetilde{\zeta} \gamma_\mu \zeta) \left(-\frac{i}{2} \varepsilon^{\rho\sigma\mu} f_{\rho\sigma} - i \partial^\mu \sigma + V^\mu \sigma \right) & = 0\, ,\label{eq:GaugeBPS-3}\\
    \frac{i}{2v} (D + \sigma H) + (\zeta \gamma_\mu \widetilde{\zeta}) \left(-\frac{i}{2} \varepsilon^{\rho\sigma\mu} f_{\rho\sigma} + i \partial^\mu \sigma + V^\mu \sigma \right) & = 0\, .\label{eq:GaugeBPS-4}
  \end{align}
  Then
  \begin{displaymath}
    \widetilde{\zeta} \gamma_1 \zeta = \zeta \gamma_1 \widetilde{\zeta} \neq 0
  \end{displaymath}
  implies that
  \begin{equation}
    \partial_1 \sigma = 0\, .
  \end{equation}
  Therefore, we can conclude that
  \begin{equation}
    \partial_\mu \sigma = 0\, ,
  \end{equation}
  i.e., $\sigma$ is constant. In the above, we prove this condition in a special frame \eqref{eq:workingFrame} \eqref{eq:workingFrame-2}, but actually the equations Eqs.~\eqref{eq:GaugeBPS-1} $\sim$ \eqref{eq:GaugeBPS-4} are frame-independent, i.e., they are valid for an arbitrary frame. Hence, in general Eq.~\eqref{eq:GaugeBPS-1} and Eq.~\eqref{eq:GaugeBPS-2} imply that
  \begin{equation}\label{eq:GaugeBPS-5}
    -\frac{i}{2} \varepsilon^{\rho\sigma\mu} f_{\rho\sigma} + V^\mu \sigma = 0\, ,
  \end{equation}
  which leads to
    \begin{equation}
      a_\mu = - \sigma C_\mu + a_\mu^{(0)}\, ,
    \end{equation}
    where $a_\mu^{(0)}$ is a flat connection, and $C_\mu$ is an Abelian gauge field satisfying
    \begin{equation}
       V^\mu = -i \varepsilon^{\mu\nu\rho} \partial_\nu C_\rho\, .
    \end{equation}
    Under the condition \eqref{eq:GaugeBPS-5}, Eq.~\eqref{eq:GaugeBPS-3} and Eq.~\eqref{eq:GaugeBPS-4} give us
    \begin{equation}
      D = -\sigma H\, .
    \end{equation}
We do not want fermionic background, hence the classical solutions of fermions are zero. To summarize, the classical solutions to the BPS equations in our model are
    \begin{equation}
      a_\mu = - \sigma C_\mu + a_\mu^{(0)}\, ,\quad \partial_\mu \sigma = 0\, ,\quad D = -\sigma H\, ,\quad \text{all other fields} = 0\, .
    \end{equation}

\section{Derivation of Some Important Relations}
  In this appendix, we prove a few crucial relations in our calculations. The relations include the second equation of Eq.~\eqref{eq:orthogonalityCond}, Eq.~\eqref{eq:crucialRel} and Eq.~\eqref{eq:defB}.

  To prove the second equation of Eq.~\eqref{eq:orthogonalityCond}, we first observe that
  \begin{equation}
    \langle S_2^* S_2^c \Phi_1,\, \Phi_2\rangle = \langle \Phi_1,\, S_2^{c*} S_2 \Phi_2\rangle\, .
  \end{equation}
  Hence, we only need to prove
  \begin{displaymath}
    S_2^* S_2^c \Phi = 0\, .
  \end{displaymath}
  Plugging in the definitions \eqref{eq:defOperators}, we can find the relation above by a long but direct calculation. In the intermediate steps we made use of the following relations:
  \begin{equation}
    \zeta^\dagger \zeta^\dagger = 0\, ;
  \end{equation}
  \begin{equation}
    V_\mu (\zeta^\dagger \gamma^\mu \zeta^\dagger) = 0\quad \textrm{since} \quad V_2 = V_3 = 0\,\, \textrm{and} \,\, \zeta^\dagger \gamma_1 \zeta^\dagger = 0\, ;
  \end{equation}
  \begin{equation}
    D_\mu \zeta^\dagger = -\frac{1}{2} H \gamma_\mu \zeta^\dagger + \frac{i}{2} V_\mu \zeta^\dagger + \frac{1}{2} \varepsilon_{\mu\nu\rho} V^\nu \gamma^\rho \zeta^\dagger\, ;
  \end{equation}
  \begin{equation}
    \gamma^\mu D_\mu \zeta^\dagger = -\frac{3}{2} H \zeta^\dagger - \frac{i}{2} \gamma^\mu V_\mu \zeta^\dagger\, ;
  \end{equation}
  \begin{equation}
    \varepsilon^{\mu\nu\rho} (\zeta^\dagger \gamma_\rho \zeta^\dagger) V_{\mu\nu} = \varepsilon^{\mu\nu\rho} (\zeta^\dagger \gamma_\rho \zeta^\dagger) A_{\mu\nu} = 0\, ,
  \end{equation}
  where
  \begin{equation}
    A_{\mu\nu} \equiv \nabla_\mu A_\nu - \nabla_\nu A_\mu\, ,\quad V_{\mu\nu} \equiv \nabla_\mu V_\nu - \nabla_\nu V_\mu\, .
  \end{equation}

  Next, to prove Eq.~\eqref{eq:crucialRel}, we just calculate
  \begin{displaymath}
    \widetilde{\Psi} (S_2 S_1^* + S_2^c S_1^{c*}) \Psi
  \end{displaymath}
  and
  \begin{displaymath}
    \widetilde{\Psi} (S_1 S_2^* + S_1^c S_2^{c*}) \Psi\, ,
  \end{displaymath}
  and then compare them to figure out their difference. The results are
  \begin{align}
    \widetilde{\Psi} (S_2 S_1^* + S_2^c S_1^{c*}) \Psi = & \,\, -i \left(z - q\sigma - (r - \frac{1}{2}) H \right) e^{-2 \textrm{Im} \Theta} \Omega (\widetilde{\Psi} P_+ \Psi) \nonumber\\
    {} & \,\, -i \left(\bar{z} - q\bar{\sigma} + (r - \frac{3}{2}) H \right) e^{-2 \textrm{Im} \Theta} \Omega (\widetilde{\Psi} P_- \Psi) \nonumber\\
    {} & \,\, + \frac{1}{2} e^{-2 \textrm{Im} \Theta} \Omega V_1 (\widetilde{\Psi} \Psi) + i (\zeta^\dagger \gamma^\mu D_\mu \Psi) (\widetilde{\Psi} \zeta) - i (\zeta \gamma^\mu D_\mu \Psi) (\widetilde{\Psi} \zeta^\dagger)\, ,
  \end{align}
  \begin{align}
    \widetilde{\Psi} (S_1 S_2^* + S_1^c S_2^{c*}) \Psi = & \,\, i \left(\bar{z} - q \bar{\sigma} + (r - \frac{1}{2}) H \right) e^{-2 \textrm{Im} \Theta} \Omega (\widetilde{\Psi} P_+ \Psi) \nonumber\\
    {} & \,\, + i \left(z - q\sigma - (r - \frac{3}{2})\right) e^{-2 \textrm{Im} \Theta} \Omega (\widetilde{\Psi} P_- \Psi) \nonumber\\
    {} & \,\, + \frac{1}{2} e^{-2 \textrm{Im} \Theta} \Omega V_1 (\widetilde{\Psi} \Psi) + i (\zeta^\dagger \gamma^\mu D_\mu \Psi) (\widetilde{\Psi} \zeta) - i (\zeta \gamma^\mu D_\mu \Psi) (\widetilde{\Psi} \zeta^\dagger)\, ,
  \end{align}
  where
  \begin{equation}
    P_\pm \equiv \frac{1}{2} (1 \pm \gamma_1)\, .
  \end{equation}
  Hence,
  \begin{align}
    (S_2 S_1^* + S_2^c S_1^{c*}) - (S_1 S_2^* + S_1^c S_2^{c*}) = & \,\, -2i \textrm{Re} (z - q\sigma) e^{-2 \textrm{Im} \Theta} \Omega P_+ - 2i \textrm{Re} (z - q\sigma) e^{-2 \textrm{Im} \Theta} \Omega P_- \nonumber\\
    = & \,\, -2i \textrm{Re} (z - q\sigma) e^{-2 \textrm{Im} \Theta} \Omega\, ,
  \end{align}
  which proves Eq.~\eqref{eq:crucialRel}.

  Finally, let us show how to prove Eq.~\eqref{eq:defB}. From Eq.~\eqref{eq:DeltalambdaLambda} we see that
  \begin{displaymath}
    (iM + \sigma \alpha \Omega) \Lambda = \Omega (-\gamma^\mu D_\mu + \frac{1}{2} H - \frac{i}{2} V_1 - i V_\mu \gamma^\mu) \Lambda\, .
  \end{displaymath}
  Using this relation, one can gradually prove that
  \begin{align}
    {} & \,\, \Omega d(\widetilde{\zeta} \Lambda) + (iM + \sigma \alpha \Omega) (\widetilde{\zeta} \gamma_\mu \Lambda) d\xi^\mu \nonumber\\
    = & \,\, \Omega H (\widetilde{\zeta} \gamma_\mu \Lambda) d\xi^\mu - \frac{i}{2} \Omega V_1 (\widetilde{\zeta} \gamma_\mu \Lambda) d\xi^\mu - \frac{i}{2} \Omega V_\mu (\widetilde{\zeta} \Lambda) d\xi^\mu \nonumber\\
    {} & \,\, + \frac{1}{2} \Omega \varepsilon_{\mu\nu\rho} V^\nu (\widetilde{\zeta} \gamma^\rho \Lambda) d\xi^\mu - i \varepsilon_{\mu\nu\rho} \Omega (\widetilde{\zeta} \gamma^\rho D^\nu \Lambda) d\xi^\mu\, . \label{eq:AppendixTemp-1}
  \end{align}
  On the other hand, there is
  \begin{equation}
    -i * \left(D(\widetilde{\zeta} \gamma_\mu \Lambda) d\xi^\mu \right) = -i \varepsilon_{\mu\nu\rho} \Omega D^\nu (\widetilde{\zeta} \gamma^\rho \Lambda) d\xi^\mu\, .
  \end{equation}
  After some steps it becomes
  \begin{equation}\label{eq:AppendixTemp-2}
    -i \varepsilon_{\mu\nu\rho} \Omega D^\nu (\widetilde{\zeta} \gamma^\rho \Lambda) d\xi^\mu = \Omega H (\widetilde{\zeta} \gamma_\mu \Lambda) d\xi^\mu - i\Omega V_\mu (\widetilde{\zeta} \Lambda) d\xi^\mu - i \varepsilon_{\mu\nu\rho} \Omega (\widetilde{\zeta} \gamma^\rho D^\nu \Lambda) d\xi^\mu\, .
  \end{equation}
  Then we only need to prove that the expressions in Eq.~\eqref{eq:AppendixTemp-1} and Eq.~\eqref{eq:AppendixTemp-2} are equal, or equivalently their difference vanishes, i.e.,
  \begin{equation}\label{eq:AppendixTemp-3}
    -\frac{i}{2} \Omega V_1 (\widetilde{\zeta} \gamma_\mu \Lambda) d\xi^\mu + \frac{i}{2} \Omega V_1 (\widetilde{\zeta} \Lambda) d\xi^1 - \frac{1}{2} \Omega \varepsilon_{1\mu\nu} V^1 (\widetilde{\zeta} \gamma^\nu \Lambda) d\xi^\mu = 0\, ,
  \end{equation}
  where we have used the fact that $V_m$ has only the $1$-component non-vanishing. The last expression can be checked explicitly by using
  \begin{align}
    {} & \quad \widetilde{\zeta} \propto (1\quad 0) \nonumber\\
    \Rightarrow & \quad \widetilde{\zeta} \gamma_1 = \widetilde{\zeta}\, ,\quad i\widetilde{\zeta} \gamma_2 = -\widetilde{\zeta} \gamma_3\, ,\quad i \widetilde{\zeta} \gamma_3 = \widetilde{\zeta} \gamma_2\, ,
  \end{align}
  where recall our convention
  \begin{displaymath}
    \gamma_1 = \sigma_3\, ,\quad \gamma_2 = -\sigma_1\, ,\quad \gamma_3 = -\sigma_2\, .
  \end{displaymath}

\bibliographystyle{h-physrev3}
\bibliography{3D}

\end{document}